\documentclass[fleqn,usenatbib]{mnras}

\usepackage{newtxtext,newtxmath}
 
\usepackage[T1]{fontenc}

\DeclareRobustCommand{\VAN}[3]{#2}
\let\VANthebibliography\thebibliography
\def\thebibliography{\DeclareRobustCommand{\VAN}[3]{##3}\VANthebibliography}

\usepackage{graphicx}	
\usepackage{amsmath}	
\usepackage{amssymb}	
\usepackage[version=4]{mhchem}

\begingroup
\catcode`\_=\active
\gdef_#1{\ensuremath{\sb{\mathrm{#1}}}}
\endgroup
\mathcode`\_=\string"8000
\catcode`\_=12

\newcommand{\sub}[1]{_{\scriptsize{\mbox{#1}}}}      
\newcommand{\supscr}[1]{^{\scriptsize{\mbox{#1}}}}       

\newcommand{\cfhtwo}[1]{\Lambda_{\scriptsize{\mbox{H}_2{#1}}}}  

\newcommand{\cm}[1]{cm$^{#1}$}                                  
\newcommand{\htwo}{H$_2$}                            
\newcommand{\hcc}{~H~\cm{-3}}                        
\newcommand{\msun}{~\mbox{M}$_{\odot}$}                      
\newcommand{\pc}{~\mbox{pc}}                         
\newcommand{\pch}{~\mbox{pc/h}}                      
\newcommand{\mpch}{~\mbox{Mpc/h}}                      

\newcommand{\jj}{\,J$_{21}$}                         
\newcommand{\jlw}{\,J$_{\mbox{\scriptsize{LW}},21}$}    
\newcommand{\jxz}{\,J$_{\mbox{\scriptsize{X0}},21}$}   

\newcommand{\nh}{n_{\mbox{\scriptsize{H}}}}        

\newcommand{\phisec}[2]{\langle \Phi\supscr{#1} (E_0\supscr{#2},\xe)\rangle} 

\newcommand{\ramses}{\texttt{RAMSES}}                
\newcommand{\ramsesrt}{\texttt{RAMSES-RT}}           
\newcommand{\xe}{x_{\scriptsize{e}}}                   

\def\hide#1{}

\title[First stars in an X-ray background]{Population~III Star Formation in an X-ray background: I.\\ Critical Halo Mass of Formation and Total Mass in Stars}

\author[J. Park, M. Ricotti and K. Sugimura]{
Jongwon Park,$^{1}$
Massimo Ricotti,$^{1}$
and Kazuyuki Sugimura$^{1,2}$
\\
$^{1}$Department of Astronomy, University of Maryland, College Park, MD 20742, USA\\
$^{2}$Astronomical Institute, Graduate School of Science, Tohoku University, Aoba, Sendai 980-8578, Japan\\
}

\date{Accepted XXX. Received YYY; in original form ZZZ}

\pubyear{2021}

\begin{document}
\label{firstpage}
\pagerange{\pageref{firstpage}--\pageref{lastpage}}
\maketitle

\begin{abstract}
The first luminous objects forming in the universe produce radiation backgrounds in the FUV and X-ray bands that affect the formation of Population~III stars. Using a grid of cosmological hydrodynamics zoom-in simulations, we explore the impact of the Lyman-Warner (LW) and X-ray radiation backgrounds on the critical dark matter halo mass for Population~III star formation and the total mass in stars per halo. We find that the LW radiation background lowers the \htwo\ fraction and delays the formation of the Population~III stars. On the other hand, X-ray irradiation anticipates the redshift of collapse and reduces the critical halo mass, unless the X-ray background is too strong and gas heating shuts down gas collapse into the halos and prevents star formation. Therefore, an X-ray background can increase the number of dark matter halos forming Population~III stars by about a factor of ten, but the total mass in stars forming in each halo is reduced. This is because X-ray radiation increases the molecular fraction and lowers the minimum temperature of the collapsing gas (or equivalently the mass of the quasi-hydrostatic core) and therefore slows down the accretion of the gas onto the central protostar.
\end{abstract}





\section{Introduction}
The first stars (or Pop~III stars) have an important role for the formation of the first galaxies and black holes. They produce the heavy elements required for the formation of the second-generation stars \citep{greif2010, Abe2021}, but their formation rate is uncertain because it is self-regulated by complex chemical and radiative feedback loops \citep{RicottiGS:2002b, wise2008}. For instance, the radiation backgrounds they produce affect their formation rate on cosmological scales \citep[][hearafter R16]{ricotti2016}. 
There has been significant progress in detecting high-z galaxies \citep[e.g.][]{bouwens2019}, but our understanding of their formation is still limited to the brightest galaxies at redshifts below $z \sim 6-10$. With James Webb Space Telescope (JWST), however, we will be able to look further and learn more on the formation processes of the first objects. For this reason, a better understanding of the formation of Pop~III star and their impact on second-generation stars is timely. The formation of the first intermediate mass black hole (IMBHs) remnants of Pop~III stars is also poorly understood because regulated by similar feedback processes as Pop~III stars \citep{regan2020}, but the IMBHs census can now be better constrained by the LIGO \citep{abbott2016} gravitational wave detectors.

\begin{figure*}
    \centering
	\includegraphics[width=0.95\textwidth]{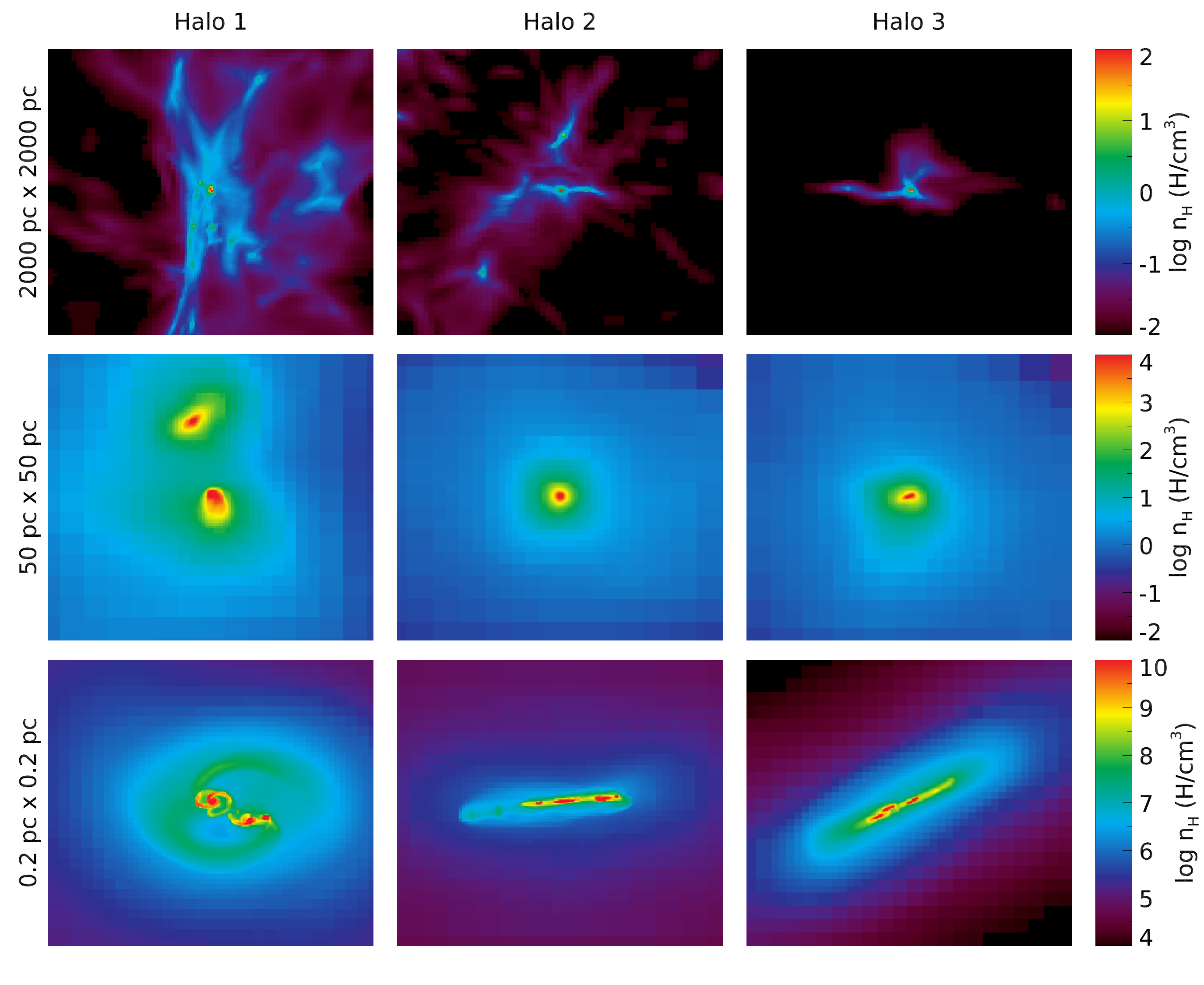}
    \caption{Snapshots of the gas density in Halo~1, Halo~1 and Halo~3 (columns from left to right, respectively) for the case without radiation backgrounds. Each row shows the gas density at three different scales (see label on the left).}
    \label{fig:halo2zoom}
\end{figure*}

The first Pop~III stars form in dark matter (DM) haloes of $\sim 10^6$ \msun\ (called minihaloes) at $z \sim 30$ \citep{tegmark1997}. In the early Universe, there are no heavy elements that can cool the gas in virialized minihaloes with $T_{vir}<10^4$~K (corresponding roughly to minihalo masses $<10^8$~\msun). Therefore, the formation of Pop~III stars relies on molecular hydrogen (\htwo) formation via the H$^-$ catalyst. This implies that the radiation regulating the amount of \htwo\ is crucial to their formation. For instance, FUV radiation in the Lyman-Werner bands (LW, $11.2 - 13.6$ eV) emitted by Pop~III stars dissociates \htwo\ \citep{omukai2001}, suppressing gas cooling in minihaloes. The mean free path of LW radiation in the intergalactic medium (IGM) is roughly $\sim 150$ times smaller than the particle horizon \citep[][R16]{RicottiGS:2001}, and much greater than the mean free path of ionizing UV radiation ($\geq 13.6$ eV). Therefore, LW photons can travel far and build a radiation background that dominates the radiation from local sources, and it is able to delay or suppress the formation of Pop~III stars \citep{haiman2000, regan2020}.
Another radiation background that affects the \htwo\ fraction is the X-ray radiation \citep[$\gtrsim 0.2$ keV,][]{venkatesan2001,xu2016}. The IGM is relatively transparent to photons in this energy range. An X-ray radiation background partially ionizes the pristine gas and increases the amount of \htwo\ that forms via $\ce{H + H^{-} -> H2 + e^{-}}$ channel. In the early Universe, high-mass X-ray binaries (HMXBs), accreating IMBHs, or supernova/hypernova explosions of Pop~III stars are possible sources of an X-ray radiation background \citep[][R16]{jeon2014,xu2016}.

The role of the X-ray background in regulating the formation of Pop~III stars is still debated. Using zoom-in simulations, \citet{jeon2014} studied the effect of X-ray radiation emitted by local sources and found negligible effects on the Pop~III star formation rate. On the other hand, through a self-consistent modelling of sources and the background they produce on IGM scales, R16 found that X-ray emission from the first sources has a positive-feedback on the number of Pop~III stars because of the enhanced \htwo\ formation and cooling. However, when the mean X-ray emission per source is above a critical value, the number of Pop~III stars per comoving volume is reduced because X-ray heating of the IGM becomes the dominant feedback effect. At the critical X-ray luminosity per source, the number of Pop~III stars per comoving \mpch$^3$\ at $z \sim 15$ is $\sim 400$, that is far larger than $\sim 10$ Pop~III stars found in the same volume without X-rays (i.e., with only LW radiation background). The limit of this model is that it considers global feedback loops but neglects local feedback effects and galaxy-scale gas dynamics.

The initial mass function (IMF) of Pop~III stars is also an important open question in Pop~III star formation theory, and has been studied numerically by many authors \citep{hirano2014,susa2014,stacy2016}. Using a number of radiative hydrodynamics (RHD)  simulations, \citet{hirano2014} and \citet[][hearafter HR15]{hirano2015} explored the final masses of Pop~III stars, taking into account photodissociation of \htwo\ by LW background, but neglecting the effect of X-ray radiation. Simulating a large sample of minihaloes, they found correlations between the final mass of Pop~III stars and properties of gas cloud or host halo, such as the gas accretion rate. \citet[][hereafter HM15]{hummel2015} explored the role of an X-ray background in determining the number of Pop~III stars and their final masses.

In this work, the first in a series, we investigate how the X-ray and LW radiation backgrounds affect the formation of Pop~III stars using a set of zoom-in cosmological simulations of minihaloes. First, we study the impact of the backgrounds on the redshift of collapse and therefore the critical mass of the dark matter minihalo when Pop~III stars form. This is tightly related to the number of Pop~III stars and pair-instability supernovae (PISNs) that will be able to be detected with JWST and the Nancy Grace Roman space telescope \citep{whalen2014}. Furthermore, we explore how the total mass in Pop~III stars depends on the intensity of the radiation backgrounds. In a companion paper \citep[][hereafter, Paper~II]{ParkRS:21b}, we study the effect of X-ray irradiation on the properties and fragmentation of protostellar discs, the multiplicity, mass function and separation of Pop~III stars.  

The paper is organized as follows. In Section~\ref{sec:sim} we introduce our simulations and methods. In Section~\ref{sec:cm} and \ref{sec:mass} we discuss how a radiation backgrounds affects the minimum mass of minihaloes forming Pop~III stars, and the final masses of Pop~III stars, respectively. In Section~\ref{sec:disc} we provide a discussion and a summary.


\section{Simulations}
\label{sec:sim}
We use the RHD \ramsesrt\ cosmological code \citep{rosdahl2013}. Its original version \ramses\ \citep{teyssier2002} is an N-body + hydrodynamic code using an adaptive mesh refinement (AMR) technique. In order to simulate the growth of haloes, gas collapse and development of discs from cosmological initial conditions, we perform zoom-in simulations. First, we run two DM-only simulations of 1 and 2 Mpc/h$^3$ boxes. We select two haloes (Halo~1 and Halo~3) from the 1 Mpc/h$^3$ box and one halo (Halo~2) from the other. Halo~1 is at the centre of a group while Halo~2 and Halo~3 are in sparsely populated regions.

Inside each zoom-in region, the mass of DM particles is $\sim 800$ \msun. The cells within the zoom regions are refined if they contain at least 8 DM particles or if their Jeans lengths are not resolved with at least $N_{J}$ cells. For the latter condition, a widely adopted value is $N_{J}=4$ \citep{truelove1997}, but we adopt the value of $N_{J}$ adaptively. At small-scales where the circumstellar disc fragments and stars form (cell size smaller than $\sim 1 \pch$, comoving), we adopt $N_{J}=16$ in order to prevent any possible artificial fragmentation and better resolve possible turbulent motions. If the size of a cell is greater than $\sim 30 \pch$ (comoving), we adopt $N_{J}=4$ to save computational time. Any cells between these two scales are refined with $N_{J}=8$. The size of a smallest cell is $0.00375 \pch$ (comoving). At z=20, this corresponds to a physical size of $2.63 \times 10^{-4} \pc$ (or $54$ au). The corresponding AMR levels are shown in Table~\ref{tab:sim}. The initial conditions of the DM-only and zoom-in simulations are generated with MUSIC \citep{hahn2011}. The assumed cosmological parameters are $h=0.674, \Omega_{m}=0.315, \Omega_{\Lambda}=0.685, \Omega_{b}=0.0493, \sigma_8=0.811$ and $n_s=0.965$ \citep{planck2018}.

\begin{table}
	\centering
	\caption{Summary of the simulations.}
	\label{tab:sim}
	\begin{tabular}{ | r | c | c | c | c | }
		\hline
        & $M_{\mbox{\scriptsize{Vir}}}$ (z=15.7) & $M_{\mbox{\scriptsize{DM}}}$ (zoom-in) & Box size & $l_{\mbox{\scriptsize{max}}}$ \\
		\hline
        Halo~1 & $7.9\times10^6$ \msun & 800 \msun & 1 Mpc/h$^3$ & 28 \\
		\hline
        Halo~2 & $4.4\times10^6$ \msun & 800 \msun & 2 Mpc/h$^3$ & 29 \\
		\hline
        Halo~3 & $7.0\times10^5$ \msun & 800 \msun & 1 Mpc/h$^3$ & 28 \\
        \hline
	\end{tabular}
\end{table}

\begin{figure}
    \centering
	\includegraphics[width=0.48\textwidth]{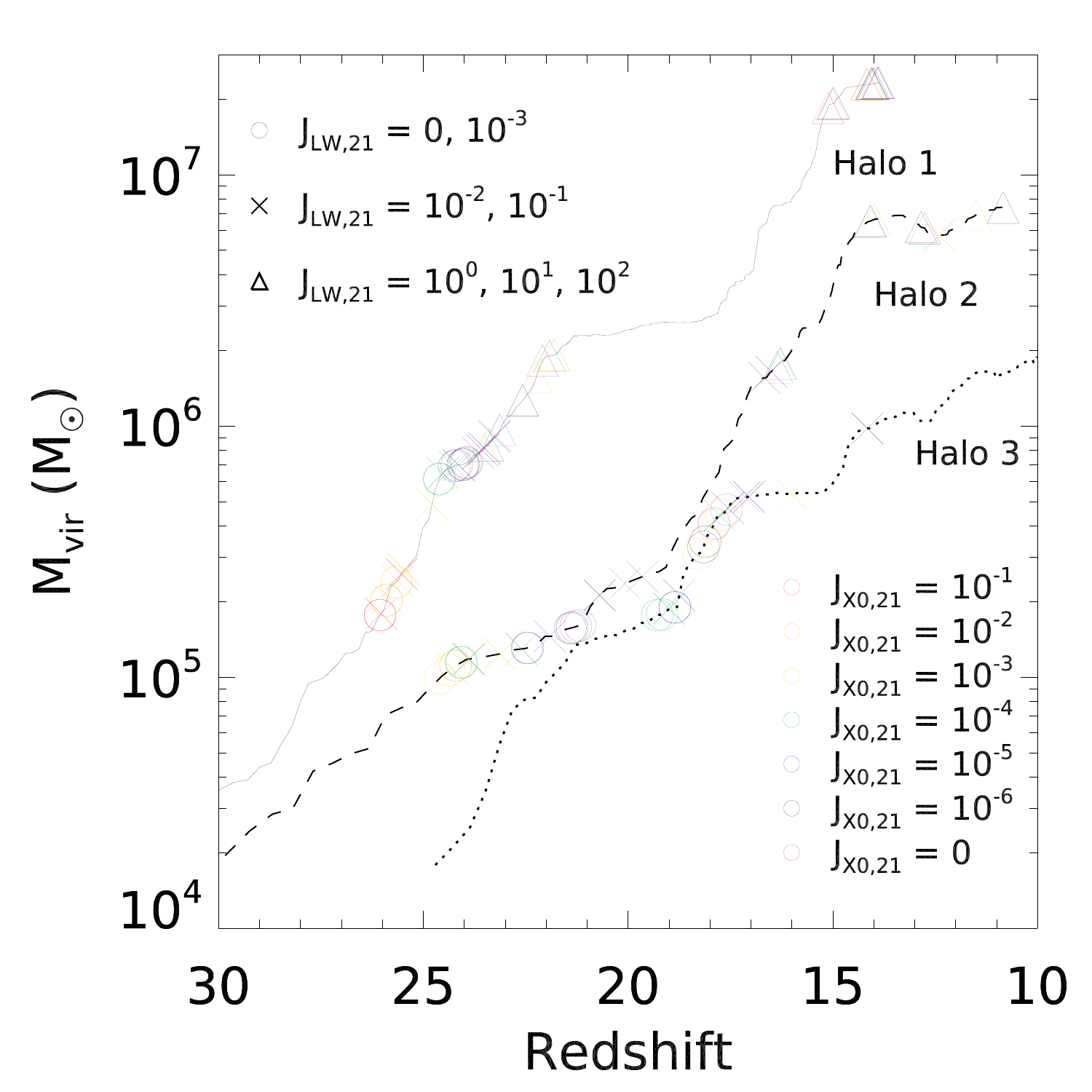}
    \caption{Halo mass as a function of redshift. The virial masses of Halo~1, Halo~2 and Halo~3 are shown with different lines. The positions of symbols refer to the redshift of the formation of Pop~III stars and the masses of their host minihaloes. Different symbols and colors indicate the intensity of LW and X-ray backgrounds as indicate by the legend (see Section~\ref{sec:bg}).}
    \label{fig:halo_growth}
\end{figure}

Snapshots of the haloes are presented in Figure~\ref{fig:halo2zoom} and a summary of the three simulations is given in Table~\ref{tab:sim}. In Figure~\ref{fig:halo_growth} we plot the virial masses of the haloes as a function of redshift. Halo~1 is more massive than the other two, and its mass reaches $\sim 2 \times 10^7$\msun\ at $z \sim 15$. Halo~2 and Halo~3 end up being haloes of $\sim 7 \times 10^6$ and $2 \times 10^6$\msun\ at $z \sim 10$. In a strong LW radiation background the collapse of the gas is delayed to when the mass of the minihalo is $\gtrsim 10^7$\msun\, close to the critical mass at which Ly~$\alpha$ cooling becomes dominant and approaching a regime thought to be the main formation channel of direct-collapse-black holes \citep{regan2020}. Excluding this extreme case, the formation of Pop~III stars in our simulations occurs when the masses of the minihaloes are between $10^5$\msun\ to $\sim 2 \times 10^6$\msun. In this range of the masses of the three haloes grow in different ways. Halo~1 grows rapidly between redshift $z \sim 30$ to $z\sim 22$ and Pop~III star formation happens during this time. Halo~3 also grows rapidly from $z \sim 19$ to $z\sim 17$ and the redshift of star formation in many simulations fall in this range. On the other hand, the mass of Halo~2 increases slowly, by a factor of two from $z \sim 25$ and $20$, and most of the star formation happens during this time. 

\subsection{Chemistry}
\label{sec:chem}

\ramsesrt\ incorporates the chemistry of hydrogen/helium ions and molecular hydrogen. In this study, the formation and destruction of \htwo\ are treated following \citet{katz2017} with some modifications. Our revised model is accurate up to gas densities $\nh \sim 10^{12}$\hcc. Improvements and validation of the primordial chemistry/cooling model with respect to the original \ramses\ version are shown in Appendix~\ref{app:chem}. The H$_2$ formation rate is,
\begin{equation}
    \frac{dx\sub{\htwo}}{dt} = -C_{coll} x\sub{\htwo} - k_{photo} x\sub{\htwo} + R n_{H I},
\end{equation}
where $x\sub{\htwo}$ is the \htwo\ fraction, $n_{H I}$ is the H I number density, $C_{coll}$ is the collisional dissociation rate, $k_{photo}$ is the photodissociation rate and $R$ is the production rate. The term $C_{coll}$ is the sum of the collisional dissociation rates of \htwo\ colliding with H, \htwo, e$^-$, He and H$^+$ \citep{glover2008,glover2010}. Its definition is
\begin{equation}
    C_{coll} = \sum_{\mbox{i}}^{} k_{coll,i} n_{i} 
\end{equation}
where i means one of five chemical species ($\ce{H, He, e^-, H^+}$ and \htwo). The term $k_{coll,i}$ is the collisional dissociation rate for species i and $n_{i}$ is its number density. The photodissociation rate is,
\begin{equation}
    k_{photo} = 4 \pi \int_{11.2 \scriptsize{\mbox{eV}}/h_{P}}^{13.6 \scriptsize{\mbox{eV}}/h_{P}} \frac{\mbox{J}_\nu}{h_{P}\nu} \sigma_{LW} d\nu.
\end{equation}
$h_{P}$ is the Planck constant, J$_\nu$ is the intensity and $\sigma_{LW}$ is the effective cross-section,
\begin{equation}
    \label{eq:shielding}
    \sigma_{LW} = 2.47 \times 10^{-18} f_{shd} ~\mbox{\cm{2}},
\end{equation}
where $f_{shd}$ is the shielding factor. To compute $f_{shd}$, we adopt the formulae in \citet{wolcottgreen2019}:
\begin{equation}
    \begin{split}
        f_{shd} &= \frac{0.965}{(1+x/b_5)^{\alpha(n,T)}}\\
        &+ \frac{0.035}{(1+x)^{0.5}} \times \exp{[-8.5 \times 10^{-4} (1+x)^{0.5}]},
    \end{split}
\end{equation}
where
\begin{equation}
    \begin{split}
        & \alpha(n,T) = A_1(T) \times \exp{(-0.2856 \times \log(n/\mbox{cm}^{-3}))} + A_2(T), \\
        & A_1(T) = 0.8711 \times \log(T/\mbox{K}) - 1.928, \\
        & A_2(T) = -0.9639 \times \log(T/\mbox{K}) + 3.892.
    \end{split}
\end{equation}
Here, $x \equiv N\sub{\htwo}/5\times10^{14}$\cm{-2}, where $N\sub{\htwo}$ is the \htwo\ column density. One difference with respect to the formula in \citet{gnedin2009} is the doppler broadening factor $b_5 = b/10^5$~cm~s$^{-1}$. For each cell, the value of b is computed with $b=(b_{turb}^2+b_{thermal}^2)^{1/2}$, where $b_{turb}=7.1 \times 10^5$~cm~s$^{-1}$ is the turbulent broadening factor \citep{krumholz2012} and $b_{thermal} = (2k_B T/m\sub{\htwo})^{1/2}$ is the thermal broadening factor \citep{richings2014}. For the latter, $m\sub{\htwo}$ is the mass of an \htwo\ molecule, $T$ is the temperature of the cell, and $k_{B}$ is the Boltzmann constant. The rate of production of \htwo\ is,
\begin{equation}
    R = \frac{k_1 k_2 \xe}{k_2 + k_5 x_{H II} + k_{13} \xe} +
    (R_{3,1} n_{H I} x_{H I} + R_{3,2} n\sub{\htwo} x_{H I}),
\label{eq:R}
\end{equation}
where $x_{H II}$ is the ionized H fraction, $\xe$ is the electron fraction and $n\sub{\htwo}$ is the \htwo\ number density. The first term on the right-hand side of the equation is the production rate through the H$^-$ channel. In this work, contrary to the other species, we do not track the out-of-equilibrium H$^-$ abundance, instead we calculate the equilibrium value using the four reaction rates of H$^-$ formation [$k_1, k_2, k_5$ and $k_{13}$ from \citep{glover2010}] as in \citet{katz2017}. At densities higher than $10^8$ \hcc, the three-body interaction (the second term in equation~\ref{eq:R}) is the dominant channel of \htwo\ formation. $R_{3,1}$ is the production rate from the reaction $\ce{3H -> H2 + H}$ \citep{forrey2013} and $R_{3,2}$ is that from $\ce{2H + H2 -> 2H2}$ \citep{palla1983}.

We make use of the results of \citet{glover2008} and \citet{glover2015} to calculate the gas cooling rate due to the collisions of \htwo\ with H, \htwo, e$^-$, He and H$^+$. The low-density limit follows equation~(37) and Table 8 of \citet{glover2008} and we adopt equation~(30) of \citet{glover2015} to calculate the LTE limit. The rate of cooling by \htwo\ is,
\begin{equation}
    \cfhtwo{} = \frac{\cfhtwo{,\mbox{LTE}}}
                {1+\cfhtwo{,\mbox{LTE}}/\cfhtwo{,n\rightarrow0}},
\end{equation}
where $\cfhtwo{,\mbox{LTE}}$ is the LTE limit and $\cfhtwo{,n\rightarrow0}$ is the low density limit.

At high-density, \htwo\ line emission is trapped by the optically thick gas, so the cooling rate is affected by the escape probability of the radiation. The cooling rate is multiplied by the following factor.
\begin{equation}
    \Bar{f}_{esc} = \frac{1}{(1+N\sub{\htwo}/N_{c})^\alpha}.
\end{equation}
The equation and parameters ($N_{c}$ and $\alpha$) are defined in \citet{fukushima2018}.

\hide{
\begin{table}
	\centering
	\caption{Summary of the radiation backgrounds.}
	\label{tab:background}
	\begin{tabular}{ | r | l | }
		\hline
        LW energy & 11.2~eV - 13.6~eV \\
		\hline
        X-ray energy & 0.2~keV - 2.0~keV \\
        \hline
        \jlw\ & $0, 10^{-3}, 10^{-2}, 10^{-1}, 10^{0}, 10^{1}, 10^{2} $ \\
        \hline
        \jxz\ & $0, 10^{-6}, 10^{-5}, 10^{-4}, 10^{-3}, 10^{-2}, 10^{-1} $ \\
        \hline
        Number of backgrounds & 49 (7 LW \& 7 X-ray) \\
        \hline
	\end{tabular}
\end{table}
}

\subsection{Radiation Background and Secondary Photo-electrons from X-ray ionization}
\label{sec:bg}

In \ramses, the ionization and heating by a radiation background can be calculated for redshifts $z < 15$ by reading a table of background spectra at different redshifts \citep{haardt2012}. In this work, however, we are interested in the radiation background in the earlier Universe at $z \sim 30-15$, hence we must follow a different approach.

\begin{figure}
    \centering
	\includegraphics[width=0.48\textwidth]{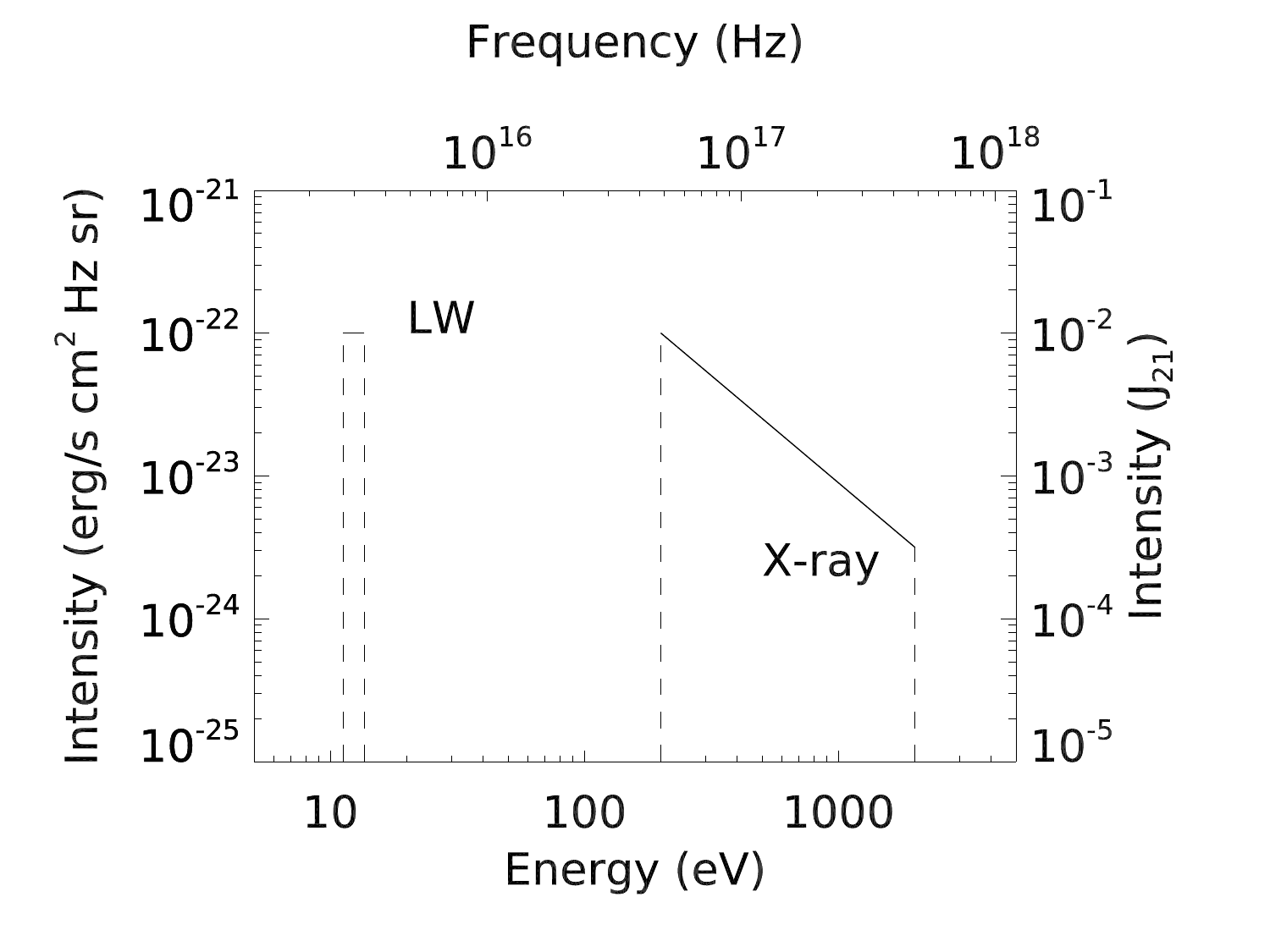}
    \caption{Spectra of LW and X-ray radiation background. The intensity in cgs unit and \jj\ are shown. The intensity of the LW background is constant and that of the X-ray background is proportional to $\nu^{-1.5}$ \citep{inayoshi2011}. We assume that ionizing UV photons (13.6~eV - 0.2~keV) cannot build a radiation background so the intensity in this range is zero. In this example, \jlw\ $= 10^{-2}$ and \jxz\ $=10^{-2}$.}
    \label{fig:background}
\end{figure}

The evolution of the radiation background and Pop~III star formation rate that produces it, are regulated by feedback loops acting on cosmological scale (see, R16). In this study, however, we focus on the formation of Pop~III stars in a single halo using zoom-in simulations. This prevents us from tracking the number of X-ray sources that contribute to the background. For this reason, we here use a simple approach. We run a large grid of simulations for each halo, with different radiation background models. Each of the models consists of a LW ($11.2 - 13.6$~eV) band background, and a soft X-ray ($0.2 - 2.0$~keV) background. Other studies \citep[HM15;][]{jeon2014,xu2016} focus on the role of HMXBs which emit harder X-ray photons ($1 - 10$~keV), but here we limit our interest to soft X-rays, mostly produced by accreating IMBHs and supernova/hypernovae exposions and remnants (R16). We assume that the specific intensity of the LW background (J$_{LW,21}$, in units of $10^{-21}$~erg~s$^{-1}$~\cm{-2}~Hz$^{-1}$~sr$^{-1}$) is constant with frequency between $11.2 - 13.6$~eV, while the X-ray background is a power law with slope $1.5$: J$_{X,21}=$\jxz$(E/E_0)^{-1.5}$, where \jxz\ is the intensity at $E_0=200$~eV \citep{inayoshi2011}. The intensity of the background is constant as a function of time in physical units (not-comoving). We explore a grid of $7\times7=49$ different combinations of LW and X-ray backgrounds: \jlw$=0,10^{-3}, 0.01,0.1,1,10,100$, \jxz$=0,10^{-6},10^{-5}, 10^{-4}, 10^{-3},0.01,0.1$. A sketch of the spectra is shown in Figure~\ref{fig:background}.
\begin{figure*}
    \centering
	\includegraphics[width=0.85\textwidth]{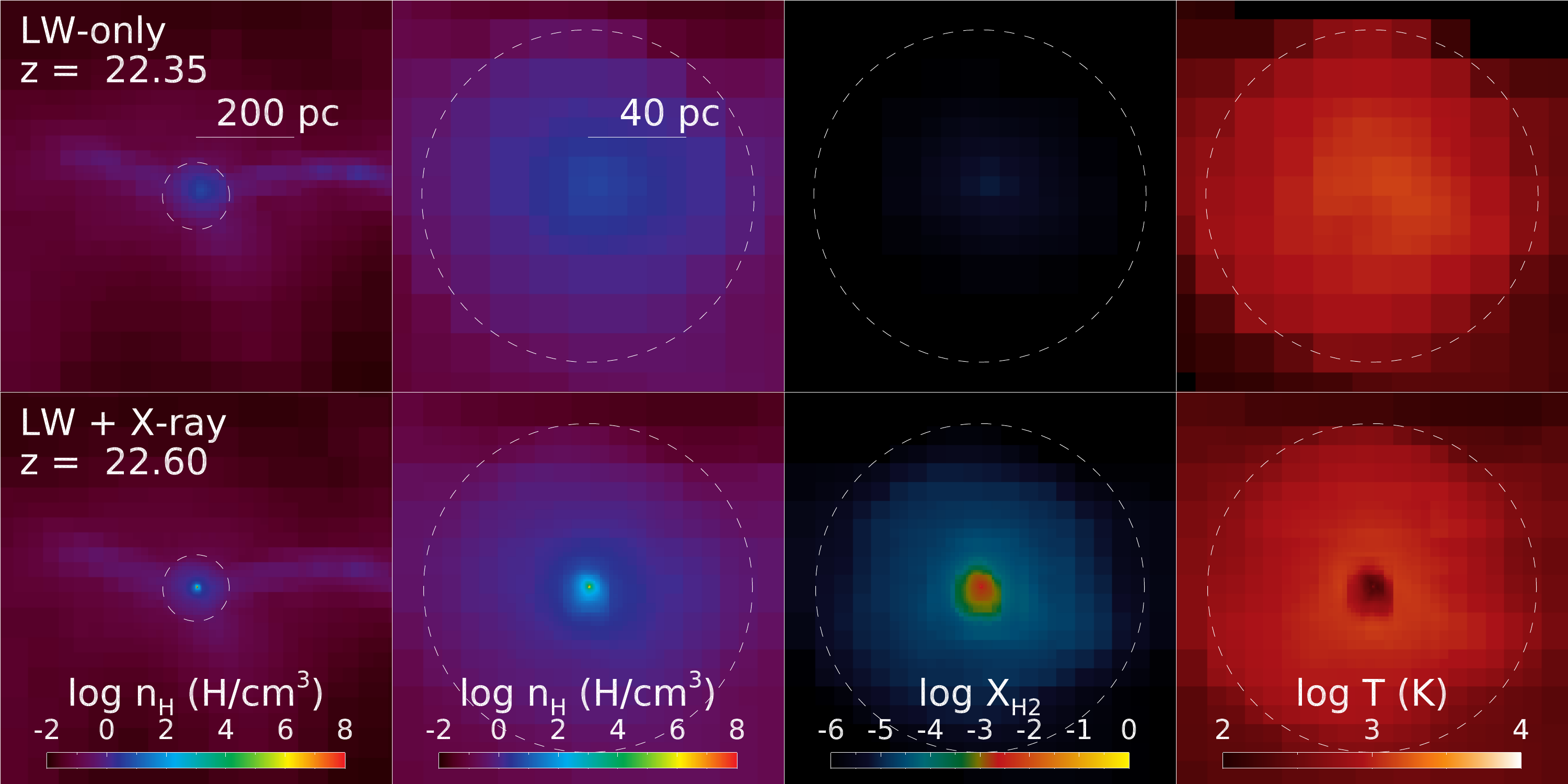}
    \caption{Halo~2 in a LW-only simulation (the top panels) and one including X-ray effect (the bottom panels). The first two columns show the hydrogen number density at different scales (shown in the top panels). The third and forth panels show the \htwo\ fraction and gas temperature, respectively (the scale is the same as in the second panel from the left). The intensity of the LW radiation is the same in both simulations (\jlw$=10^{-1}$). The intensities of X-ray backgrounds are \jxz$=0$ and $10^{-4}$, respectively. The white circle indicates the virial radius of the halo.}
    \label{fig:map_example}
\end{figure*}

If a high-energy X-ray photon ionizes a neutral hydrogen, the resultant photo-electron also has large kinetic energy. This electron may produce secondary ionizations by colliding with other neutral hydrogen atoms or be thermalized and heat the gas by colliding with other electrons. We model the secondary ionization/heating following \citet{shull1985} and  \citet{RicottiGS:2002a}. The ionization rate of species $i$ is:
\begin{equation}
    \begin{split}
        -\frac{d x_{i}}{dt} & = x_{i} \zeta \supscr{i} + \sum_{\scriptsize{\mbox{j=H I, HeI}}} \frac{n_{j}}{n_{i,tot}} \zeta \supscr{j} \phisec{i}{j}, \\
        & = (1 + f_{ion,i}) \zeta \supscr{i},
    \end{split}
\end{equation}
where $f_{ion,i}$ is the secondary ionization fraction of species i, $E_0\supscr{i} = h_{P}\nu-I_{i}$ is the photo-electron energy and $\phisec{i}{j}$ is the average number of secondary ionization per primary electron of species j \citep[see][]{RicottiGS:2002a}. $I_{i}$ is the ionization potential and $\zeta \supscr{i}$ is the photo-ionization rate
\begin{equation}
    \zeta_{i} = 4 \pi \int_{I_{i}/h_{P}}^{\infty} \frac{\mbox{J}_\nu}{h_{P}\nu} ~\sigma_{i} ~d\nu,
\end{equation}
where $\sigma_{i}$ is the ionization cross-section of i \citep{verner1996}. The rate of heating due to species i is
\begin{equation}
    \begin{split}
        \frac{de}{dt} &= 4 \pi \int_{I_{i}/h_{P}}^{\infty} \frac{\mbox{J}\nu}{h_{P}\nu} ~\sigma_{i} E_{h}(E_0\supscr{i},x_{e}) ~d\nu \\
        &= f_{heat,i} \Gamma_{i}.
    \end{split}
\end{equation}
where $e$ is the internal energy. A fraction of the energy of photo-electrons is used to ionize the gas and therefore the heating of the gas is less efficient \citep{RicottiGS:2002a}. This effect is considered with the factor $E_{h}(E_0\supscr{i},x_{e})$. The photoheating rate neglecting secondary photo-electron ionization is
\begin{equation}
    \Gamma_{i} = 4 \pi \int_{I_{i}/h_{P}}^{\infty} \frac{\mbox{J}\nu}{h_{P}\nu} (h_{P}\nu-I_{i}) ~\sigma_{i} ~d\nu.
\end{equation}
The secondary ionization fraction $f_{ion,i}$ is large (e.g. $f_{ion,H~I} \sim 17.3$) when the gas is almost neutral and its value converges to zero with increasing $\xe$. On the contrary, the heating fraction $f_{heat,i}$ is low in a neutral gas (close to 0.1) and converges to 1 with increasing $\xe$. This can be understood as in a neutral gas (low $\xe$), most of the photo-electrons collide with neutral hydrogen and ionize the gas, while in a highly ionized gas, photo-electrons collide with other electrons and thermalize.

\subsection{Tracking Clumps}
\label{sec:track}

We impose that the Jeans length is resolved with at least $N_{J}=16$ cells up to the maximum AMR level near the halo centre. Once the maximum level is reached, however, we cannot refine them further. In order to prevent artificial fragmentation at the maximum refinement level caused by a decreasing Jeans length, we suppress the cooling of cells with the maximum refinement level following the method in \citet{hosokawa2016}. This is done by multiplying the cooling function by the factor
\begin{equation}
    \label{eq:climit}
    C_{limit} = \exp{ \left[ -\left( \frac{\xi-1}{0.1} \right)^2 \right]} \hspace{0.5cm} \mbox{(if $\xi > 1$)},
\end{equation}
where $\xi = f_{limit} (\Delta x/\lambda_{J})$ with the cell size $\Delta x$ and Jeans length $\lambda_{J}$. We assume $f_{limit}=12$ as in \citet{hosokawa2016}.

To investigate the growths of protostars and the initial mass function, many studies employ sink particles/cells and calculate the gas accretion onto and radiative feedback from them (e.g. HR15; S20). In this study, we do not use sink particles, hence we neglect accretion and feedback physics associated with them. Our goal is to perform many simulations for various radiation backgrounds to capture the effects of X-rays on the collapse and fragmentation. Instead of using sink particles, we flag cells with $C_{limit} < 10^{-4}$ in each run. Using a friends-of-friends algorithm, we link neighboring cells together to identify dense clumps. Each clump represents a quasi-hydrostatic core that would collapse to form a star if we did not impose limitations to cooling rate and the resolution. Throughout this article, we use each clump as a proxy for a single Pop~III star.
\begin{figure*}
    \centering
	\includegraphics[width=0.95\textwidth]{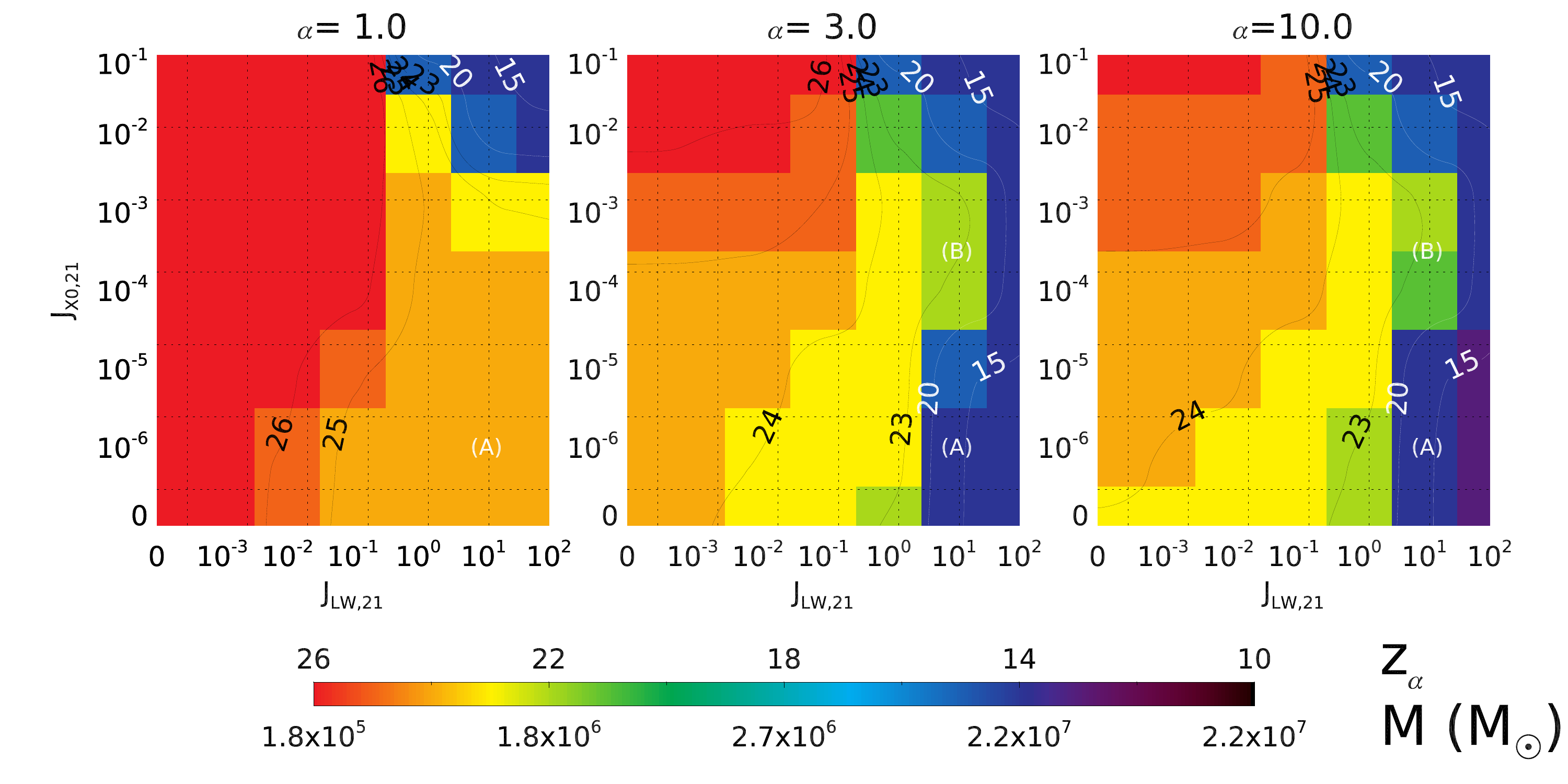}
    \caption{Redshift of collapse to density $10^\alpha$ ($z_\alpha$) for Halo~1 as a function of \jlw\ and \jxz. From left to right, $\alpha$ is 1, 3 and 10. The smoothed iso-contours for $z_\alpha= 15$, 20, 23, 24, 25 and 26 are drawn on the top of the color map. Under the color bar, the mass of the halo at particular redshift is shown. The locations in each panel marked by letter-labels are used for a more detailed explanations of results in the text.}
    \label{fig:z_halo1}
\end{figure*}
\section{Results: I. Critical Minihalo Mass for Pop~III formation}
\label{sec:cm}

In a gas of primordial composition, formation of \htwo\ is inefficient in minihaloes below a critical mass due to the low electron fraction produced by collisional ioniation during virialization \citep{tegmark1997}. However, pristine gas can cool more efficiently, even in small mass minihaloes, if an X-ray radiation background enhances the electron fraction.
Figure~\ref{fig:map_example} demonstrates the effectiveness of the X-ray background in promoting Pop~III star formation. When the halo is irradiated by a constant (as a function of time) LW background (top panels), the maximum gas density in the minihalo at redshift $z=22.6$ is below $10$\hcc, and the gas collapses to form a dense core at $z \sim 16.5$, when the mass of the minihalo is $1.6 \times 10^6$\msun. If the halo is exposed to the same LW background but also an X-ray background (bottom panels), the gas at the halo centre cools efficiently (the right panels shows the temperature) thanks to the enhanced \htwo\ fraction (see the third panel from the left). The cooling enables the formation of a dense core at $z=22.6$, when the halo mass is $\sim 1.5 \times 10^5$\msun, a critical mass about ten times lower than the case without X-ray irradiation.

To quantify the effect of the X-ray background we define the critical mass as the mass of a minihalo at the redshift when the Pop~III star forms. To this end, we define the parameter $z_\alpha$, that is the redshift at which the central gas density reaches $10^\alpha$\hcc. For example, if the central gas density reaches $10^1$\hcc\ at $z=25$ and $10^8$\hcc\ at $z=23$, $z_1=25$ and $z_8=23$, respectively. In this section, we assume that Pop~III stars form at $z_{10}$. From the time of formation of the protostellar core, it takes roughly $ 10^4 - 10^5$ years for feedback to halt gas accretion and reach the final Pop~III star mass (S20). This time-scale is much shorter than the typical variations of the formation time caused by different intensities of the LW or X-ray radiation backgrounds. The later phases of evolution of the gas ($\nh >10^{10}$\hcc), when a protostellar disc is formed, are discussed later in Sections~\ref{sec:mass} and Paper~II.

The three panels in Figure~\ref{fig:z_halo1} show $z_1$, $z_3$ and $z_{10}$, respectively, for Halo~1 for the $7\times7$ grid of radiation backgrounds. The middle and right panels are nearly identical, meaning that once the gas reaches a density above $10^3$~\hcc, for the range of backgrounds considered here, the collapse cannot be halted and proceeds rapidly to densities $\sim 10^{10}$~\hcc. 
The panel on the left shows that the gas density reaches $10$\hcc\ by $z \sim 24$ for any value of the LW background, unless the X-ray intensity is larger than \jxz$>10^{-3}$ and \jlw$>10$. This is because the core gas density of Halo~1 at virialization is $\sim 10$\hcc\ even without any core contraction due to gas cooling \cite[see][for the calculation of the core gas density assuming an isothermal equation of state and $T \approx T_{vir}$]{Ricotti2009}. Therefore $z_1$ is always reached by Halo1 unless the X-ray background is strong enough (\jxz$>10^{-3}$) to heat the IGM to $T_{IGM} > T_{vir}$ and cooling from \htwo\ formation (that is increased by X-ray photo-ionization) does not offset significantly the X-ray heating (\jlw$>10$).

Let's now focus on the right panel, showing the redshift of Pop~III star formation. In a zero background (the bottom left corner), the Pop~III star forms at $z \sim 24$ when the halo mass is $\sim 7 \times 10^5$\msun. Increasing the LW background intensity to \jlw$\sim 10$ delays Pop~III formation to $z \sim 14$ when the halo mass exceeds $2 \times 10^7$\msun\ (label A).
\begin{figure*}
    \centering
	\includegraphics[width=0.95\textwidth]{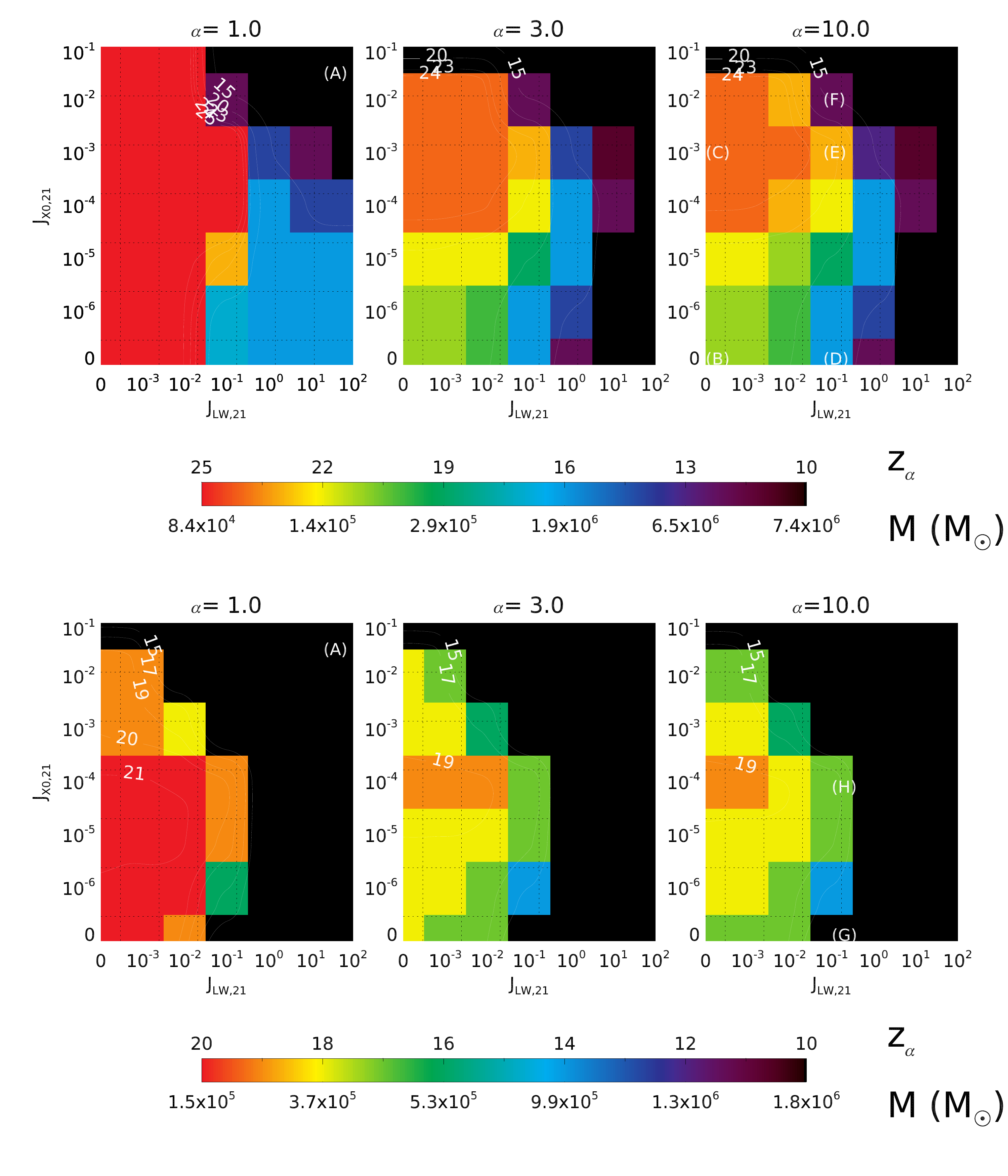}\\
    \caption{Redshift of collapse to density $10^\alpha$ ($z_\alpha$) for Halo~2 (top panels) and Halo~3 (bottom panels). The meaning of the labels and contour plots is the same as in Figure~\ref{fig:z_halo1}. Note that for Halo~3 the iso-contour lines are at $z= 15, 17, 19, 20$ and 21.}
    \label{fig:z_halo2}
\end{figure*}
\begin{figure*}
    \centering
	\includegraphics[width=0.95\textwidth]{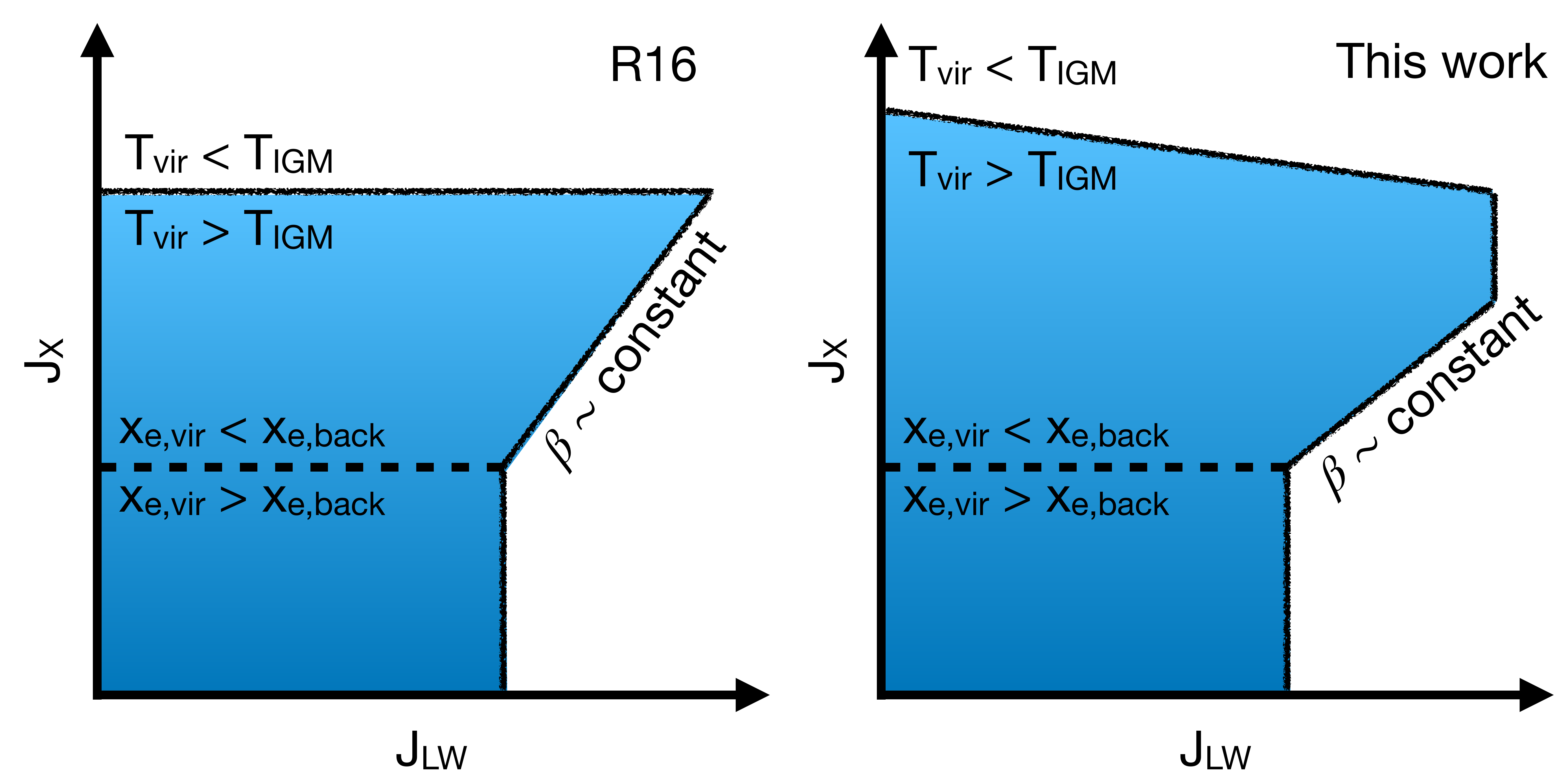}\\
    \caption{Schematic diagrams of Figure~\ref{fig:z_halo1} and \ref{fig:z_halo2}. The colored regions represent the region on the grid where the formation of Pop~III is allowed. The left panel shows the region expected by the model in R16, and the right one simplifies the trend of our simulations. $T_{vir}$ and $T_{IGM}$ indicate the temperatures of a virialized halo and the IGM, respectively. $x_{e,vir}$ is the electron fraction of the halo in the absence of an X-ray background and $x_{e,back}$ is the fraction of electrons due to additional ionization by the X-ray background radiation. The parameter $\beta$, related to the mean spectral energy distribution of the first sources of light, is the ratio of the integrated energy in the X-ray band and the LW band (see R16).}
    \label{fig:z_schematic}
\end{figure*}
The figure also shows that $z_{10}$ increases, and the critical mass decreases, with increasing X-ray intensity. This is consistent with HM15 that also find earlier onset of gas collapse in an X-ray background. We find that the X-ray background is particularly effective in offsetting the negative feedback effect of a strong LW background. In a weak X-ray background, LW intensity  \jlw$\sim 10$ delays the formation of Pop~III stars to $z \sim 14$ (label~A in the figure), but if an X-ray background \jxz$\sim 10^{-4}-10^{-3}$ is present, the central gas can collapse to form Pop~III stars at $z \sim 22$ (label~B) when the mass of the halo is below $1.8 \times 10^6$\msun. Compared to $2 \times 10^7$\msun\ at $z \sim 14$, this is a reduction of the critical mass by one order of magnitude.  In a cosmological context, a reduction of the critical mass by an order of magnitude implies that the number of minihaloes that are able to form Pop~III stars will increase by roughly the same factor.
As a side note, \cite{regan2020} found that LW radiation backgrounds with \jlw$\sim 0.1 - 10$ suppresses the formation of Pop~III stars and allows direct-collapse-black holes to form in atomic cooling haloes (with mass $>10^8$\msun). Our results suggest that a strong X-ray background may offset this effect enabling Pop~III star formation even in a strong LW background.

Figure~\ref{fig:z_halo2} shows the same results as Figure~\ref{fig:z_halo1}, but for Halo~2 and Halo~3, respectively. At all redshifts, Halo~2 is smaller in mass than Halo~1, and Halo~3 is the smallest halo (see Figure~\ref{fig:halo_growth}). The results for these two haloes are qualitatively similar to Halo~1, with some quantitative differences explained below. 
For these smaller mass haloes, when both the LW and X-ray intensity are large, the gas density fails to reach $10$\hcc\ (label~A in the figure). This is because, due to their lower masses and virial temperatures, the core densities of these haloes at virialization are $<10$\hcc, hence sufficiently rapid cooling is necessary to reach that density. A LW background \jlw$\ge 10^{-1}$ is sufficient to suppress cooling and prevent the gas from reaching $10$\hcc\ before $z \sim 16$. The collapse of gas to higher densities (from $\alpha = 3$ to $10$) follows the same trend as that of Halo~1. An increase in the LW intensity suppresses the formation of \htwo\ and delays Pop~III star formation. Compared to Halo~1, however, significant delay of Pop~III star formation occurs for a weaker LW radiation background. For instance, $z_{10}$ of Halo~1 decreases dramatically between \jlw=$1$ and $10$ while this occurs between \jlw=$0.1$ and $1$ for Halo~2 and \jlw=$0.01$ and $0.1$ for Halo~3. In a strong LW background, Pop~III stars fail to form before $z \sim 13$ ($z \sim 10$) for Halo~2 (Halo~3). 

The positive feedback by an X-ray background is also observed in Halo~2 and Halo~3. Without any LW  or X-ray radiation backgrounds $z_{10} = 21.27$ ($18$) in Halo~2 (Halo~3) and the critical mass is $\sim 1.5 \times 10^5$\msun\ (location~B). In a strong X-ray background (\jxz$=10^{-3}$, location~C), $z_{10} = 24.6$ ($\sim 31$~Myr earlier) in Halo~2 and the critical mass decreases to $9.5 \times 10^4$\msun (while it decreases by a factor of $\sim 3$ for Halo~3). When LW radiation is stronger (\jlw$\sim 0.1$), the effect of X-rays becomes more pronounced: in Halo~3 Pop~III star does not form without X-ray irradiation (location~G) and in Halo~2 the critical mass is $1.6 \times 10^6$\msun\ ($z_{10} = 16.50$, location~D). Adding X-ray radiation (location~E and H), the critical mass is lower by more than a factor of 10 for Halo~2 ($1.2 \times 10^5$\msun, $z_{10} = 23.11$) and is about $\sim 3.0 \times 10^5 - 10^6$\msun\ for Halo~3.

If the X-ray intensity increases further, however, the heating of the IGM becomes the dominant feedback mechanism, suppressing the formation of Pop~III star as in the analytic models of R16. Increasing the X-ray intensity by a factor of 10 (location F) suppresses gas collapse and Pop~III star formation does not occur until $z=12.60$ when the halo mass is $6.6 \times 10^6$\msun. If \jxz$=0.1$, the heating by a radiation background shuts down gas collapse completely and the formation of a Pop~III star is suppressed until $z = 10$.

In all the haloes the reduction of the critical minihalo mass produced by an increasing X-ray background is largest in an intense LW background. This can be readily understood because no matter how strong is the boosting of H$_2$ formation by X-rays, there is a floor to the lowest critical mass, that is dictated by the inefficient cooling rate of H$_2$ in gas with temperature below $\sim 100-200$~K. Hence, virialized gas in minihaloes with $T_{vir}<100-200$~K cannot cool rapidly even if the gas has high molecular fraction. A rough estimate of this floor critical mass of dark matter minihaloes is,
\begin{equation}
    M_{cr}(\mbox{min}) \approx 10^5~M_\odot \left(\frac{1+z_{vir}}{20}\right)^{-3/2},
\label{eq:mcr}
\end{equation}
where we have assume $T_{vir} = 200$~K. This limiting minimum mass could be significantly reduced by HD cooling, that is neglected in our work.


In order to better understand the common qualitative features described above for the three haloes, in Figure~\ref{fig:z_schematic} we show a sketch of the region on the grid in which Pop~III stars can form. The sketch on the left is based on the analytical model in R16, while the sketch on the right illustrates the results of this study based on hydrodynamical simulations.

\begin{figure*}
    \centering
	\includegraphics[width=\textwidth]{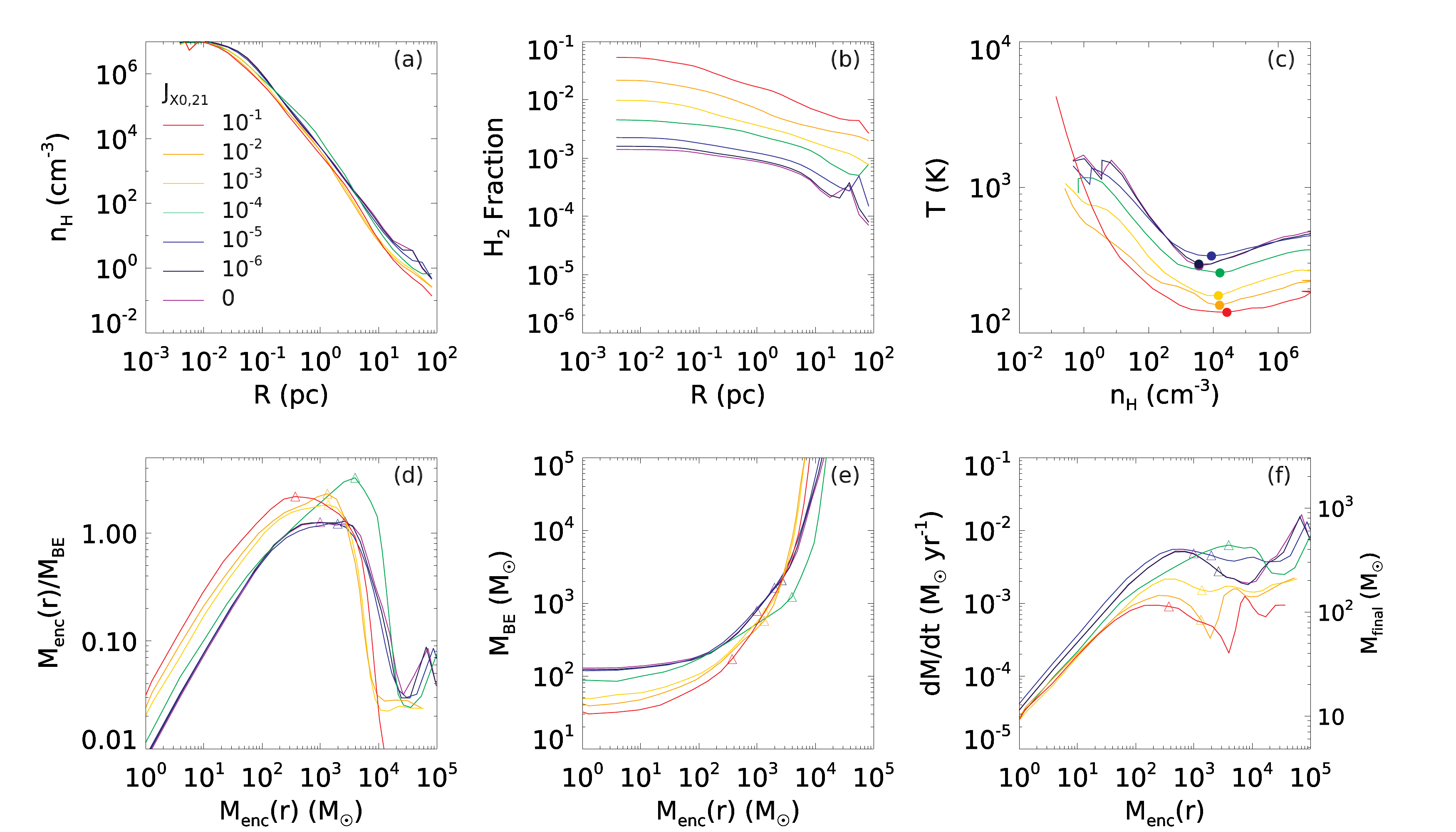}
    \caption{Properties of the gas in Halo~1 when the central density reaches $10^7$\hcc.  The X-ray intensity \jxz\ is color-coded and the LW intensity is zero. Panel (a): The gas density profile, $\nh$, as a function of the distance from the halo centre, $R$. Panel (b): The \htwo\ fraction profile as a function of $R$. Panel (c): The gas phase diagram constructed from the temperature and density profiles. Each dot represents the minimum temperature of the gas. Panel (d): The ratio of the enclosed mass to the Bonnor-Ebert mass as a function of the enclosed mass. Each peak defines the characteristic radius and it is marked by a triangle. We use the same symbols in Panel (e) and (f) to show the characteristic radii. Panel (e): The Bonnor-Ebert mass versus the enclosed mass. Panel (f): The rate of gas accretion (the left axis) and the corresponding final mass (using Equation~\ref{eq:finalmass}). The symbols indicate the accretion rate at the characteristic radii, and define the final mass of Pop~III stars.}
    \label{fig:profile_halo1}
\end{figure*}

When an X-ray background is intense (top line in the sketch), gas photo-heating is the dominant feedback process. In this case the temperature of the IGM ($T_{IGM}$) and the halo virial temperature ($T_{vir}$) determine the formation of Pop III~stars. If $T_{vir}< T_{IGM}$ the gas in the halo cannot collapse until the halo mass grows enough and $T_{vir}$ exceeds $T_{IGM}$. R16 assumed that only haloes with $T_{vir} > T_{IGM}$ can form Pop~III stars and this is represented as an horizontal line in the left panel. In the hydrodynamics simulations, however, the formation of Pop~III stars in a strong X-ray radiation shows a dependence on the intensity of the LW radiation and thus this line is slanted. We speculate that this is due to the additional cooling by \htwo\ in the absence of a strong LW background. The position of the line on the diagram is shifted upward (to higher \jxz) if the halo is more massive and has higher virial temperature. 

The lower part of the diagram can be understood considering the degree of X-ray ionization and H$_2$ formation. If the X-ray radiation is weak (below the dashed line) the fraction of electrons produced by X-ray photoionization ($x_{e,back}$) is lower than the residual electron fraction from the epoch of recombination and the electron fraction produced by collisional ionization in the virialized halo ($x_{e,vir}$). In this region the positive feedback by the X-ray background is negligible. The position of the line is mostly determined by the growth rate of the halo. Halo~1, that grows quickly and becomes more massive than the other two, has higher virial temperature and thus higher collisional ionization rate. In addition, the halo is irradiated for a shorter period of time before the formation of Pop~III stars. Therefore, the intensity of the X-ray background must be higher (\jxz$\sim 10^{-5}$ for Halo~1, see Figure~\ref{fig:z_halo1}) to increase sufficiently $x_{e,back}$ and have an influence on the critical mass of the halo. On the other hand, the position of the dashed line is lower for Halo~2 and 3 (\jxz$\sim 10^{-6}$, Figure~\ref{fig:z_halo2}) because they are irradiated by X-rays for a longer time, hence weaker X-ray background is sufficient to produce $x_{e,back}>x_{e,vir}$.

Above the dashed line the X-ray photoionization is the dominant positive feedback. In this region of the parameter space, the increase \htwo\ fraction from X-ray ionization compensates the \htwo\ dissociation by the LW background, making the formation of Pop~III stars possible in a stronger LW background. The slanted line on the right side the region shows this trend. The slope of the isocountour lines in this region is roughly linear: $J_X \propto J_{LW}$. This can be understood in the context of the R16 model in which the critical mass was found to depend on the parameter $\beta$, that is the ratio of the energies in two energy bins in the mean spectrum of the sources $\beta \equiv E_{X}/E_{LW} \sim \mbox{J}_X/\mbox{J}_{LW}$. 

\begin{figure*}
    \centering
	\includegraphics[width=0.95\textwidth]{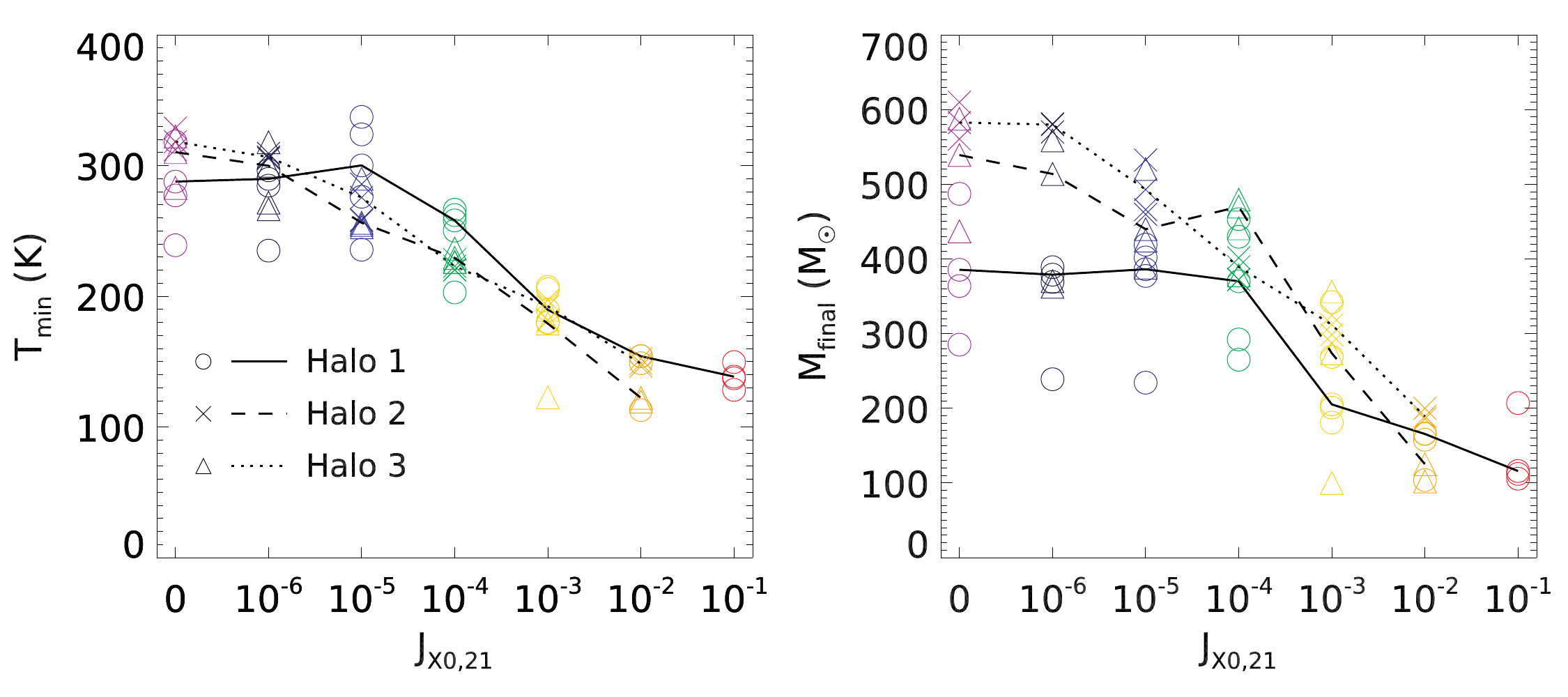}
    \caption{({\it Left}.) Minimum gas temperature, $T_{min}$, as a function of X-ray intensity \jxz. The X-ray intensity is also color-coded as in Figure~\ref{fig:profile_halo1}. We distinguish different host halos with different symbols (see legend). To illustrate the results for different halos more clearly, we plot trend lines connecting median values for each halo (different line styles as in the legend). ({\it Right}.) Same as the left panel, but showing $M_{final}$ as a function of \jxz.}
    \label{fig:TMX}
\end{figure*}

In conclusion, considering the finite sampling in the \jlw, \jxz\ plane due to the limited number of simulations in this work, our results are in good qualitative agreement with the analytic model in R16, although the simulations show that the R16 model neglects some physical processes that are important when the X-ray irradiation is strong, and/or when minihaloes masses grow rapidly. In absence of radiation backgrounds the critical halo mass we find is in agreement with R16 (about $10^6$\msun). We note that in R16 the smallest mass halo in which Pop~III stars can form is $\sim 3 \times 10^4$\msun, obtained when a strong X-ray irradiation is considered at very high redshift. In this work we find a minimum mass about three times larger ($\sim 10^5$\msun). This is in agreement with the estimate of the minimum critical mass in equation~(\ref{eq:mcr}), given that the analytical work considers also very rare haloes forming at $z>30$.

\section{Results: II. Total Mass in Pop~III Stars}
\label{sec:mass}

Our simulations follow the collapse of protostars for $5\times 10^4 - 10^5$~years after the time of formation, defined here when the core density reaches $\sim 10^{10}-10^{11}$\hcc. However, we neglect radiation feedback from the accreting protostars that can dissociate \htwo\ (FUV radiation) and produce winds powered by photo-heating from hydrogen and helium ionization. Therefore the masses of the Pop~III stars at the end of our simulations are (in most cases) overestimated and provide an upper limit to the Pop~III mass. Without radiation feedback the masses of the Pop~III stars grow at a nearly constant rate from the time of formation. Radiation feedback is expected to both reduce the accretion rate and halt the accretion after $\sim 2\times 10^4$~years (see S20).

In order to estimate the effect of radiation feedback in reducing the final masses of Pop~III stars, we use an empirical relationship based on previous work by HR15 that includes UV radiation feedback. HR15 provides a relation between the final mass of Pop~III star and the accretion rate onto the protostellar core
\begin{equation}
    M_{final} = 250 ~\mbox{M}_{\odot} \left( \frac{dM/dt|_{cr}}{2.8 \times 10^{-3} \mbox{M}_{\odot} \mbox{yr}^{-1}} \right)^{0.7},
    \label{eq:finalmass}
\end{equation}
where $M_{final}$ is the final mass and $dM/dt|_{cr}$ is the accretion rate at the characteristic radius. We define the characteristic radius consistently with the definition in HR15: When the central gas density reaches $10^7$\hcc, we compute the enclosed mass ($M_{enc}(r)$) and the Bonnor-Ebert mass \citep{ebert1955,bonnor1956} as a function of the radius, $r$:
\begin{equation}
    M_{BE}(r) \approx 1050 ~\mbox{M}_{\odot} \left( \frac{T}{200 \mbox{K}} \right)^{3/2} \left( \frac{\mu}{1.22} \right)^{-2} \left( \frac{\nh}{10^4 \mbox{cm}^{-3}} \right)^{-1/2} \left( \frac{\gamma}{1.66} \right)^2,
\end{equation}
defined as in equation~12 in \citet{hirano2014}. Here, $\mu$ is the mean molecular weight and $\gamma=5/3$ is the adiabatic index. The characteristic radius where we estimate the accretion rate is defined where $M_{enc}(r)/M_{BE}(r)$ reaches its maximum value.

\begin{figure*}
    \centering
	\includegraphics[width=0.95\textwidth]{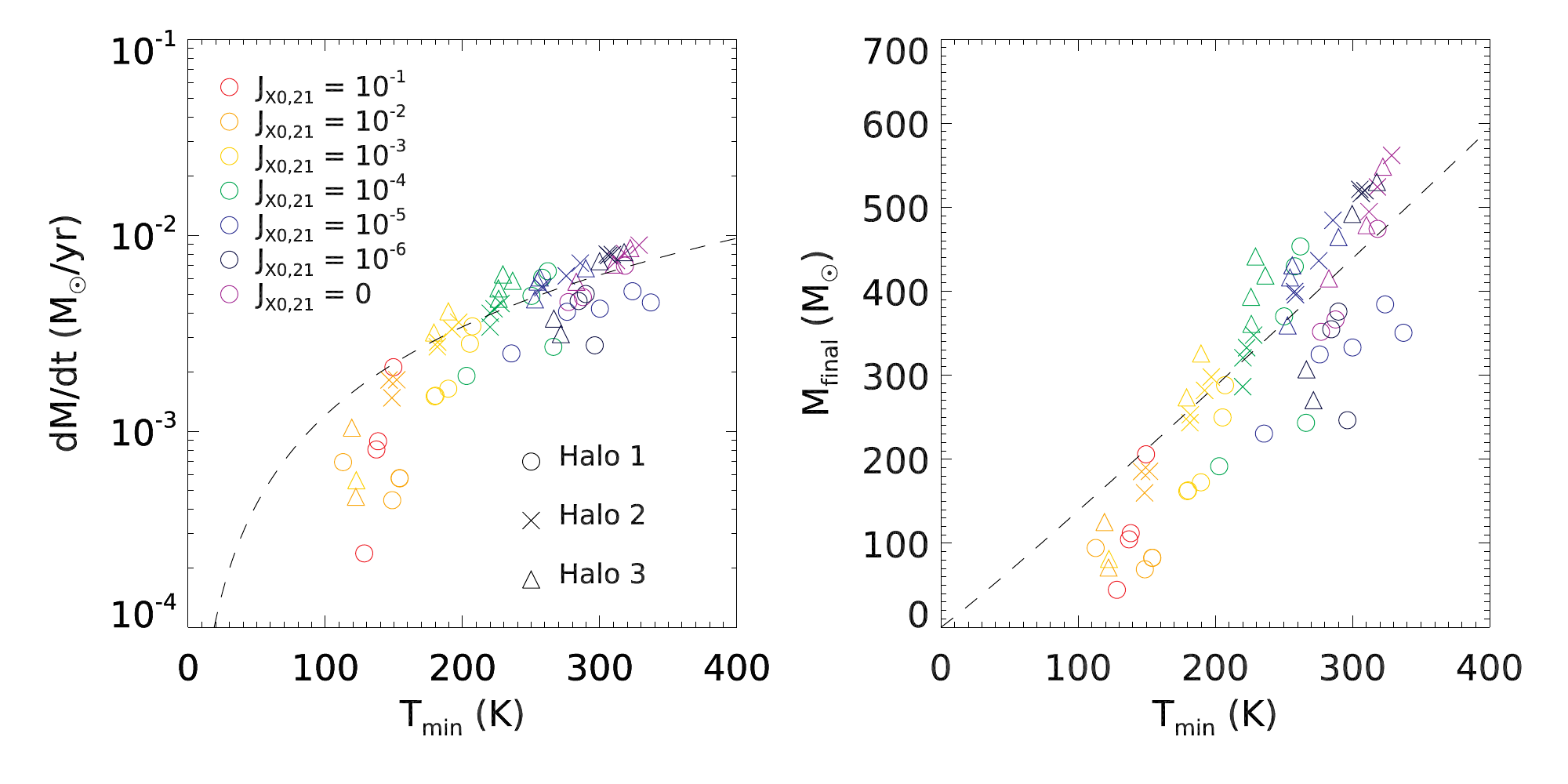}
    \caption{The accretion rate (left panel) and final mass (right panel) as a function of $T_{min}$. Different symbols indicate different halos and the X-ray intensity \jxz\ is color-coded. Simulations with critical mass $> 10^6$\msun\ are omitted due to their larger deviations from the trend. In both panels, the analytic estimations (Equations~\ref{eq:mdot_Tmin} and \ref{eq:final_Tmin}) are drawn with dashed lines.}
    \label{fig:TM}
\end{figure*}

Figure~\ref{fig:profile_halo1} shows various gas properties of Halo~1 when the central density reaches $10^7$\hcc\ for simulations with different values of \jxz\ (see legend). All the panels show spherically averaged quantities. Panel (a) shows that the gas density profile follows the self-similar solution of \citet{larson1969} and \citet{penston1969}, as shown also in \citet{omukai1998}. In Panel (b) the \htwo\ fraction profiles are shown. We can see that the \htwo\ fraction increases with the intensity of the X-ray background. This is due to the enhanced H$^-$ abundance produced by the increased fractional ionization of the gas ($x_e$) from direct X-ray ionization and secondary ionization from fast photo-electrons. Panel (c) shows the gas effective equation of state ($T$ as a function of $\nh$ for the collapsing gas obtained by matching the radial density and temperature profiles). As the X-ray intensity increases, the temperature of the gas at each given density decreases. Due to the self-similarity of the collapse, the temperature shown here also represents the temperature evolution of the gas core as a function of time as the core reaches higher densities. The temperature at the lower density is the virial temperature that depends on the halo mass and the redshift. The symbols mark the minimum temperatures ($T_{min}$), roughly reached at end of the initial free-fall phase when the core gas density is $\nh \sim 10^4$\hcc\ \citep{bromm2002}. The minimum temperature shows a correlation with the X-ray intensity: the stronger the X-ray intensity, the lower the temperature (see also the left panel of Figure~\ref{fig:TMX}). This result may appear counter intuitive as X-ray irradiation typically leads to enhanced heating rather than cooling. Instead we find that, unless the X-ray irradiation is so strong to suppress collapse and Pop~III star formation, the enhanced H$_2$ abundance and cooling rate dominates over the X-ray heating rate. This is the key physical result that will allow us to interpret most aspects of the simulations results. 

In Panel (d), we plot the ratio of the enclosed mass to the Bonnor-Ebert mass. At radii where the enclosed mass is greater than $M_{BE}$, we find a collapsing core. Vice versa, at radii larger than the collapsing core, thermal pressure supports the gas against collapse. Following \citet{hirano2014}, we define the peak value of $M(r)/M_{BE}$ as the characteristic core radius in each simulation (marked as triangles). In Panel (e), we plot the Bonnor-Ebert mass as a function of the enclosed mass. In general, for higher X-ray intensity $M _{BE}$ at the characteristic radius is smaller. Hence, X-ray irradiation reduces the the mass of the collapsing core and other properties determined by it (such as the accretion rate). Panel (f) shows the gas accretion rate and the final Pop~III star mass estimated using equation~(\ref{eq:finalmass}). Although we show $dM/dt$ at all radii, only the value of $M_{final}$ at the characteristic core radius (marked by a triangle) is meaningful. With increasing X-ray intensity, the accretion rate decreases and the core becomes smaller in mass. The expected final Pop~III mass is also lower due to the lower accretion rate. This result is consistent with HM15 in that, in an X-ray background, enhanced \htwo\ fraction provides an efficient cooling and enables smaller gas clouds to collapse.

Although the general trend with X-ray intensity is clear and monotonic, exceptions are observed. In the simulation with \jxz$=10^{-4}$ (green lines) the gas shows a higher than expected accretion rate. We find that another halo is within the virial radius of Halo~1 ($\sim 100$~pc) when the dense core forms. At this time a strong radial inflow of gas develops within a few parsec from the halo centre which accounts for the higher accretion rates seen in Figure~\ref{fig:profile_halo1}. We speculate that an interaction between two haloes triggers the strong gas inflow but we do not discuss this further as it is beyond the scope of this work.

To summarize, $T_{min}$ in panel (c) is a key property to estimate the characteristic mass of Pop~III star and $T_{min}$ is directly related to the X-ray intensity \jxz. In Figure~\ref{fig:TMX} we provide $T_{min}$ and the final Pop~III star mass $M_{final}$, as a function of the X-ray intensity \jxz. Note that the relationship between $T_{min}$ and \jxz\ depends on the assumed X-ray spectrum of the sources and the duration of the X-ray irradiation of the halo (hence the time dependence of \jxz\ and the growth rate of the minihalo mass). In this paper we have used a relatively soft power-law spectrum, but if the dominant sources of X-rays are HMXBs the spectrum is expected to be harder.
Although the three haloes are different in mass and redshift of Pop~III star formation, the figure shows the general trend of a decreasing $T_{min}$ and $M_{final}$ with increasing X-ray intensity. However, Halo~1 (solid lines) deviates from this trend in a weak X-ray background (\jxz$\lesssim 10^{-5}$), showing a roughly constant $T_{min}$. This is consistent with the discussion in Section~\ref{sec:cm}. A bigger halo has higher collisional ionization rate and is irradiated for a shorter period of time before forming Pop~III stars. For these reasons, the threshold of the positive X-ray feedback is higher (dashed lines in Figure~\ref{fig:z_schematic}). 
\begin{figure*}
    \centering
	\includegraphics[width=\textwidth]{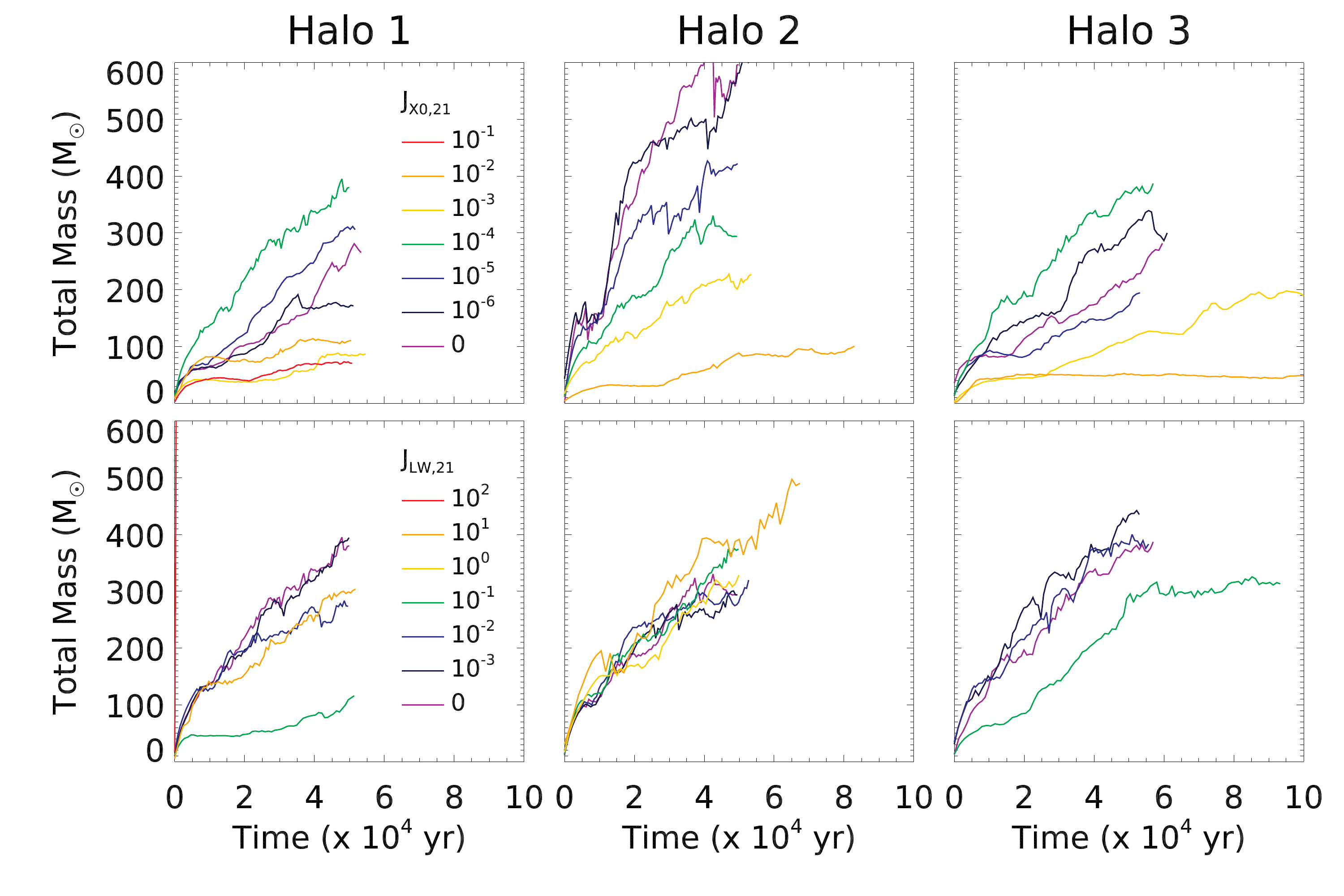}
    \caption{Evolution of the total mass of clumps since the formation of the first one in three haloes. Top panels shows the total masses in different X-ray backgrounds (\jlw$=0$). Bottom panels show those in various LW background for \jxz$=10^{-4}$. The intensity of the background is color-coded as shown in the legend.}
    \label{fig:evol_mtot}
\end{figure*}

The accretion rate at the critical radius can be estimated analytically \citep{hirano2014} assuming spherical symmetry and simple dimensional analysis: $dM/dt \sim M_J/t_{ff}$, where $M_{J}=(4\pi/3)\rho(c_s t_{ff})^3$ is the Jeans mass of the quasi-hydrostatic core at temperature $T_{min}$ and $t_{ff}=1/\sqrt{G\rho}$ is the free-fall time. 
Hence,
\begin{equation}
    \label{eq:mdot_Tmin}
    \begin{split}
        \frac{dM}{dt}  
        &\sim 
        3.4 \times 10^{-3}~\mbox{\msun\ yr}^{-1} ~\left( \frac{T_{min}}{200~\mbox{K}} \right)^{3/2} \left( \frac{\mu}{1.22} \right)^{-3/2},
    \end{split}
\end{equation}
showing that the accretion rate depends only on the gas sound speed or temperature of the collapsing core.
Since the gas is nearly neutral ($\mu = 1.22$), using equation~(\ref{eq:finalmass}) we obtain the relationship between $T_{min}$ and the final Pop~III mass:
\begin{equation}
    \label{eq:final_Tmin}
    M_{final} = 287~\mbox{M}_{\odot} ~\left( \frac{T_{min}}{200~\mbox{K}} \right)^{1.05}.
\end{equation}
We found that a strong X-ray radiation background lowers $T_{min}$ and the Jeans mass, reducing the mass of the collapsing core (HM15), thus reducing the accretion rate. 

Figure~\ref{fig:TM} shows the accretion rates at the critical radius and final mass in Pop~III stars for all the simulations forming Pop~III stars, as a function of the minimum temperature $T_{min}$. The different symbols refer to the three halos we have simulated and the colors refer to the X-ray background intensity (see legend in the the left panel of the figure). As we discuss with Figure~\ref{fig:TMX} they all follow a similar relationship: $T_{min}$ decreases with increasing X-ray intensity and this leads to a slower gas accretion and hence lower mass of the Pop~III star.
The dashed lines in the left and right panels show the result of the simple analytic model for the accretion rate (equation~\ref{eq:mdot_Tmin}) and Pop~III mass (equation~\ref{eq:final_Tmin}), respectively. The accretion rates from the simulations are consistent with this simple model, but the uncertainties in determining the location of the characteristic radii (and thus the accretion rates) produces a significant scatter with respect to the analytic calculation, especially for strong X-ray irradiation. The accretion rate $dM/dt$ and $M_{final}$ for Halo~2 (crosses) and Halo~3 (triangles) show a much tighter relationship with $T_{min}$, probably because these two haloes grow relatively slowly and do not experience major perturbations. On the other hand, Halo~1 (circles), that grows rapidly and experiences a major merger at some point, shows a larger scatter and $dM/dt$ is slightly biased to be lower than the other cases at a given temperature. 

Figure~\ref{fig:evol_mtot} shows that the total mass of the protostellar cores grows roughly linearly as a function of time. In other words, the accretion rate is roughly constant as a function of time, and its value decreases with increasing X-ray intensity similarly to $M_{final}$. The trend is quite robust, for instance, when \jxz$\geq 10^{-2}$ (orange and red lines in the top panels) the protostars grow at a significantly slower rate and the total mass in Pop~III stars is lower at any given time. Again, deviations from the general trend are observed (green lines) in Halo~1 and Halo~3. As discussed in earlier in the section, a halo interaction may cause the rapid gas accretion. When Halo~3 is irradiated by X-ray background of \jxz$=10^{-1}$ (the orange line in the upper right panel) the growth of the clump stagnates after it reaches $\sim 50$\msun. We observe that a bar-like structure develops inside the characteristic radius. Due to the increase in the accretion rate the clump at the centre grows in $\sim 10$~kyrs but does not grow as most of the disc gas is consumed. 

The bottom row in Figure~\ref{fig:evol_mtot} shows the total mass as a function of time for different intensities of the LW backgrounds (see legend) keeping fixed the X-ray intensity (\jxz$=10^{-4}$). We do not observe any dependence of the protostellar core growth rate on the LW intensity. Probably this can be explained by H$_2$ self-shielding (equation~\ref{eq:shielding}) that increases rapidly when the gas density exceeds $\sim 10^1 - 10^2$\hcc. Hence, after the formation of the protostar the LW radiation is shielded and does not play a significant role in determining $T_{min}$ and thus the final Pop~III mass. The effect of the LW background is limited to setting the redshift of Pop~III star formation. Because the protostar forms and accreates at different redshifts (for the same halo but with different radiation backgrounds), other factors such as the mass of the halo or mergers with other minihaloes are more important in determining the Pop~III star mass growth than the gas cooling physics. An extreme example of this is the case of Halo~1 irradiated by a very strong LW background (\jlw$=100$, red line in the bottom-left panel).
The gas in the halo starts condensing when the mass of the halo reaches $\sim 2 \times 10^7$\msun\ (the virial temperature $\sim 9,000$~K), when Ly~$\alpha$ atomic cooling start to become important. The accretion rate is very rapid and both $M_{final}$ and the core mass exceed $10^4$\msun. This is consistent with \citet{regan2020} in that a LW background allows black hole seeds to form by suppressing the formation of Pop~III stars in small mass minihaloes below the atomic cooling limit.
\begin{figure}
    \centering
	\includegraphics[width=0.48\textwidth]{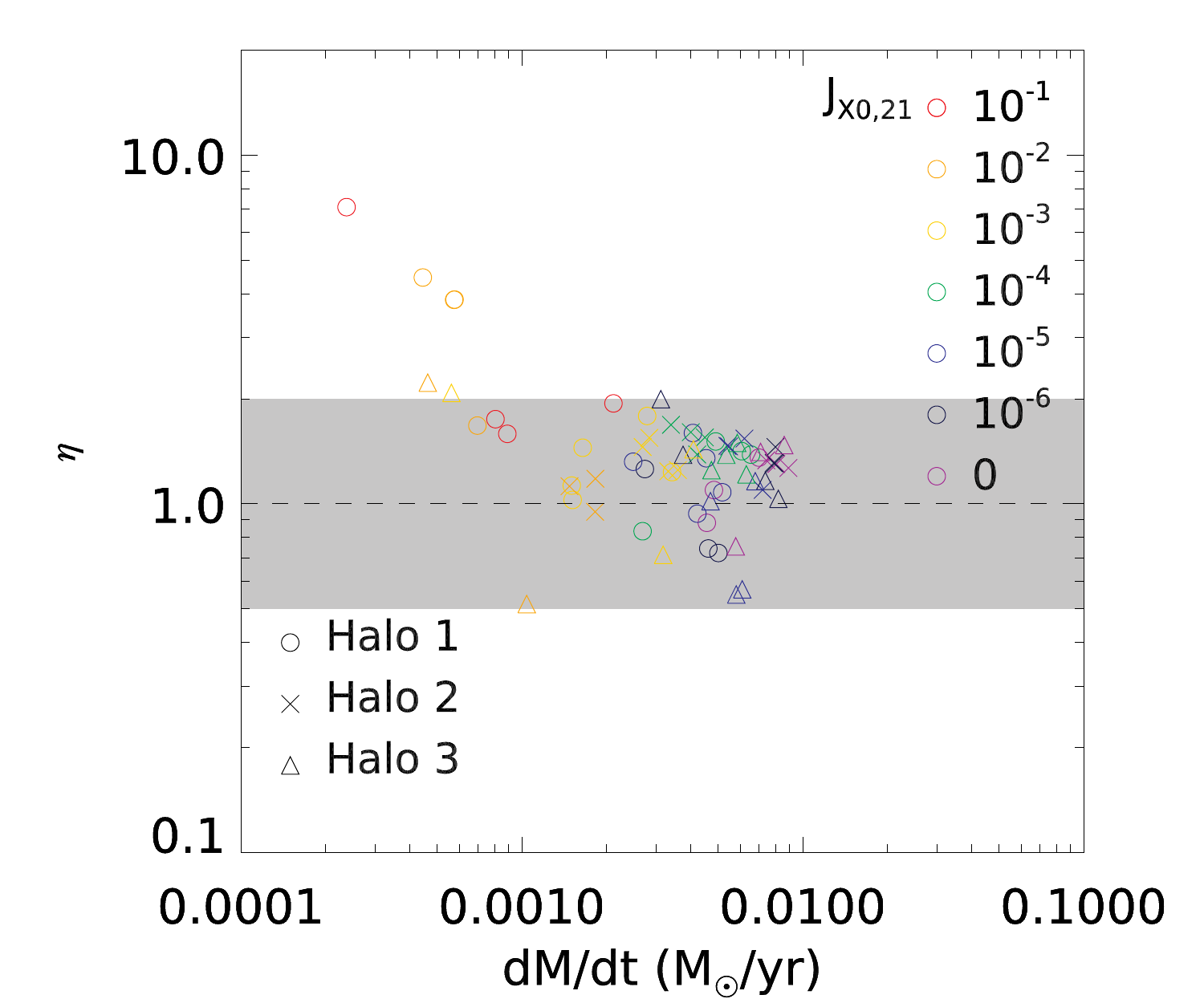}
    \caption{The dimensionless ratio $\eta$ of two accretion rates (see Equation~\ref{eq:eta}) as a function of $dM/dt|_{cr}$, defined when the central density is $10^7$\hcc. The parameter $\eta$ is the ratio of the average accretion rate on the Pop~III star ({\it i.e.}, the total mass at $t_*=5 \times 10^4$~years, divided by $t_*$)  to $dM/dt_{cr}$. Different symbols refer to different host halos and \jxz\ is color-coded. To guide the eye, the shaded region encompass a change of a factor of two around unity: $0.5 \leq \eta \leq 2.0$. Simulations with critical mass of the host halos larger than $10^6$\msun\ are omitted.}
    \label{fig:eta}
\end{figure}

Since the accretion rate onto the protostar is fairly constant, the final mass in Pop~III stars increases linearly with time. In order to determine the final mass in Pop~III stars we therefore need to estimate the typical timescale for feedback to halt the accretion. This can be derived from equation~(\ref{eq:finalmass}) as long as the accretion rate remains constant and equal to the value  
$dM/dt|_{cr}$ estimated at the critical core radius at $\nh=10^7$\hcc:
\begin{equation}
\tau_{SF} = \frac{M_{final}}{dM/dt|_{cr}} = 89~{\rm kyrs}\left( \frac{dM/dt|_{cr}}{2.8 \times 10^{-3} \mbox{M}_{\odot} \mbox{yr}^{-1}} \right)^{-0.3},
\end{equation}
In Figure~\ref{fig:eta} we show the ratio between the average growth rate of the protostar $\langle dM/dt\rangle \approx M(t_*)/t_*$, where we use $t_*=50$~kyrs for this period and the accretion rate $dM/dt|_{cr}$ at $\nh=10^7$\hcc,
\begin{equation}
    \eta = \frac{\langle dM/dt \rangle}{dM/dt|_{cr}}
    \label{eq:eta}
\end{equation}
as a function of $dM/dt|_{cr}$. The ratio $\eta$ is of the order of unity with a spread within a factor of two (shaded region) in most cases. Since these two accretion rates are nearly the same, the final mass in Pop~III stars estimated as $M(\tau_{SF})$ is the same as $M_{final}$ in equation~(\ref{eq:finalmass}).

\section{Summary and Discussion}
\label{sec:disc}

The first stars have an important effect on the formation of the first galaxies by regulating the formation of metal-enriched stars \citep{RicottiGS:2002b,RicottiGS:2008,wise2008}. 
LW and X-ray photons can travel large distances without being absorbed in the early Universe, building up a radiation background. In addition, X-rays can penetrate deep into star forming clumps producing both heating and ionization. Each photo-electron has high energy but in a mostly neutral gas fast photo-electrons deposit a large fraction of their kinetic energy into secondary ionizations \citep{shull1985}. Because of these properties, it is thought that early objects (first stars and their remnants such as SNe, PISNe, HMXBs, and IMBHs) build a LW and X-ray radiation backgrounds that self-regulates their formation \citep[][R16]{venkatesan2001,jeon2014,xu2016}. 

In this study, using zoom-in simulations of three minihaloes with different mass and irradiated by different intensities of the LW and X-ray backgrounds, we investigate the effect of X-ray/LW radiation on the formation of the first stars and on their characteristic mass. Below we summarize the key results of the simulations.
\begin{enumerate}
    \item We confirm qualitative results of previous analytic models (R16), that an X-ray radiation background promotes the initial gas collapse in small mass minihaloes, while a LW background delays it by regulating the amount of \htwo\ formation. If the X-ray background is too intense, the feedback effect by gas heating suppresses Pop~III star formation in haloes with virial temperature $T_{vir}<T_{gas}$.
    Below this threshold X-ray intensity, the increase in H$_2$ formation due to the increased ionization fraction of the gas, reduces the mass above which a minihalo can host a Pop~III star (the critical mass) to $\sim 10^5$\msun.  The positive feedback effect of X-rays is most important when it offsets the negative feedback of an intense \htwo-dissociating LW radiation background. For example, X-rays can reduce the critical mass by a factor of $\sim 2$ in a weak LW intensity, while the reduction is by a factor of ten when \jlw$=10^{-1}$. Hence, X-ray radiation can increase the number of Pop~III stars forming in the early Universe by about a factor of ten.
    
    \item X-ray irradiation produces a net cooling effect on the collapsing protostellar core by increasing the H$_2$ fraction. Efficient gas cooling reduces the gas sound speed and consequently the accretion rate on collapsing protostellar cores. Therefore the final mass in Pop~III stars is lower in X-ray irradiated minihaloes.
\end{enumerate}
The results in this work constitute a first step to understand the self-consistent evolution of the number of Pop~III stars forming in the early universe. As shown in R16 using analytic calculations, the critical minihalo mass above which Pop~III stars can form and the intensity of the radiation backgrounds are self-regulated by 
a feedback loop on cosmological scales. While this work considers only a grid of assumed values of the background, the lower value of the critical mass we derive is roughly a factor of $3$ higher than the value found in R16 ($\sim 3 \times 10^4$\msun). Although cosmological simulations are required to make solid predictions, a simple scaling argument using Press-Schechter formalism \citep{press1974,sheth1999}, suggests that the number of Pop~III stars according to our simulation results would be a factor of 3-4 lower than in R16: $\sim 100$ instead of $\sim 400$ in 1\mpch\ estimated by R16.

In addition, following R16, we assume that soft X-ray ($\sim 0.2 - 2.0$ keV) is the dominant source of ionization and our spectra do not cover harder X-ray ($\sim 2.0 - 10$ keV). Whether adding photons in this energy bin has positive or negative feedback will be studied in the future.

The model of R16 assumes that one halo has only one Pop~III star. As other simulations \citep[S20;][]{turk2009,clark2011} and this work suggest, however, the formation of multiple Pop~III stars in one halo. This implies that the prediction of a radiation background depends on the IMF of Pop~III stars. Due to their high masses, they are emitting at Eddington rate $1.25 \times 10^{38}$~ergs~s$^{-1}$~(M/M$_{\odot}$) \citep{bromm2001} so the LW intensity per halo depends on the total mass of the stars. On the contrary, if they explode as PISNe with similar energy ($\sim 10^{52}$~ergs), the X-ray intensity per halo depends on the number of Pop~III stars. Furthermore, the fraction of HMXBs \citep[$\sim 35\%$,][]{stacy2013} can be affected by the X-ray background. This suggests the number density of Pop~III stars or their explosions that JWST \citep{whalen2014} will detect may provide constraints of the IMFs and X-ray physics.

HM15 finds that the X-ray radiation background is shielded by the dense central gas and therefore plays a minor role in determining the IMF of Pop~III stars. Since the self-shielding of the X-ray background is not considered in our simulations, we perform a test run with Halo~2 to investigate if it is an important factor. In this simulation, we assume that radiation is shielded completely if the density of a cell exceeds $10^4$\hcc\ since the optical depth of X-rays becomes close to unity at this radius. This simple prescription overestimates the shielding effect since the optical depth monotonically increases with decreasing radius. The test run gives results similar to those of the original one. The redshift of Pop~III star formation is $z_{10} = 24.597$, the same as in runs neglecting self-shielding ($24.595$). In addition, the total mass in Pop~III stars at $t=50$~kyrs after the formation of the first clump are very close ($\sim 220$~\msun\ versus $240$\msun\ in the fiducial run). Finally, the multiplicity of Pop~III stars is two in the original run, while in the run with X-rays self-shielding a third small mass Pop~III star is formed. We speculate the causes of the similar results are the following. First, as discussed in Section~\ref{sec:mass}, the size of the quasi-hydrostatic core (at $\nh \sim 10^4$\hcc) determines the accretion rate and the properties of Pop~III stars. For this reason, X-ray photons do not change the later evolution even if they do not penetrate to the centre of the halo. Secondly, the collapse to a higher densities occurs on a timescale comparable to the ionization/recombination timescales. 
Hence, the dense gas retains a memory of the temperature and ionization fraction at lower densities, as shown by panel (c) of Figure~\ref{fig:profile_halo1} up to densities $\sim 10^7$~\hcc. We confirm that this behaviour continues up to densities $n_H \sim 10^{10}$~\hcc. At densities $n_H > 10^4$~\hcc, the gas temperature raises with increasing density, hence compression heating \citep[or viscous heating in discs at densities $>10^{10}$~\hcc,][]{Kimura2021} dominates over photoionization heating. As HM15 has pointed out, however, this result may be sensitive to the column density of the halo. So what matters to the X-ray shielding is the growth history of the host halo.

This work does not include two potentially important physical processes: HD cooling and radiative feedback from protostars. HD provides additional cooling at low temperatures ($\sim 100$~K) and thereby can reduce the characteristic mass of Pop~III stars \citep{yoshida2007}. Its formation rate is proportional to that of \htwo\, which is regulated by an X-ray background. Furthermore, Figure~\ref{fig:profile_halo1} and \ref{fig:TMX} show the minimum temperature is a function of X-ray. Therefore, X-ray radiation can be crucial to the onset of HD cooling. The role of X-ray radiation in HD formation has been explored by \citet{nakauchi2014}, but not tested yet with hydrodynamics simulation. Finally, a full treatment of radiation feedback from accreting protostars, rather than the semi-empirical approximation used in this work, is important for a more accurate determination of the final masses of Pop~III stars \citep[S20;][]{hosokawa2011,hosokawa2016}. Therefore, including the neglected or approximated physical processes mentioned above (X-ray self-shielding, HD cooling and radiation feedback) will be the focus of future work.

\section*{Acknowledgements}

We thank Dr. Harley Katz for sharing his version of the RAMSES code with us. All the simulations were performed with the use of Deepthought2 cluster operated by the University of Maryland (http://hpcc.umd.edu). MR acknowledges the support by NASA grant 80NSSC18K0527. KS appreciates the support by the Fellowship of the Japan Society for the Promotion of Science for Research Abroad.




\bibliographystyle{mnras}
\bibliography{mnras} 




\appendix

\section{Validation of Primordial Chemistry and Cooling in \ramses}
\label{app:chem}

\begin{figure*}
    \centering
	\includegraphics[width=0.98\textwidth]{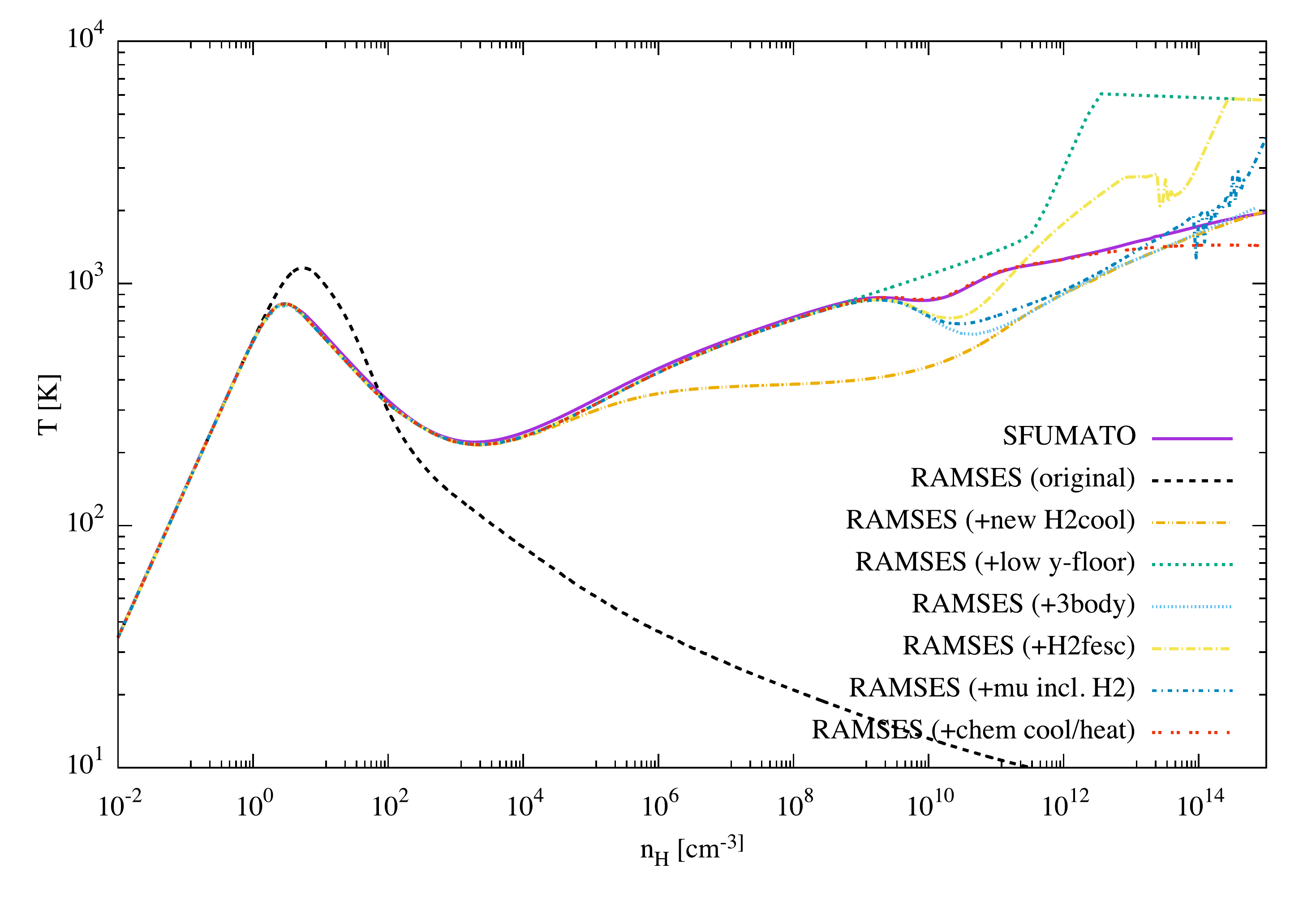}
    \caption{Phase diagram (temperature as a function of hydrogen number density) for a one-zone calculation in which compression heating balances molecular hydrogen cooling. The different lines (see legend) refer to incremental improvements of the model with respect to the  original implementation in \ramses. The validation of the result is done by comparing our implementation in \ramses\ to published results by S20 using the AMR code SFUMATO. Our results are accurate to densities $n_h \sim 10^{12}$~\hcc, but the error remains relatively small ($< 15$ per cent) up to densities $n_H \sim 10^{14}$~\hcc.}
    \label{fig:app1}
\end{figure*}

\begin{figure*}
    \centering
	\includegraphics[width=0.48\textwidth]{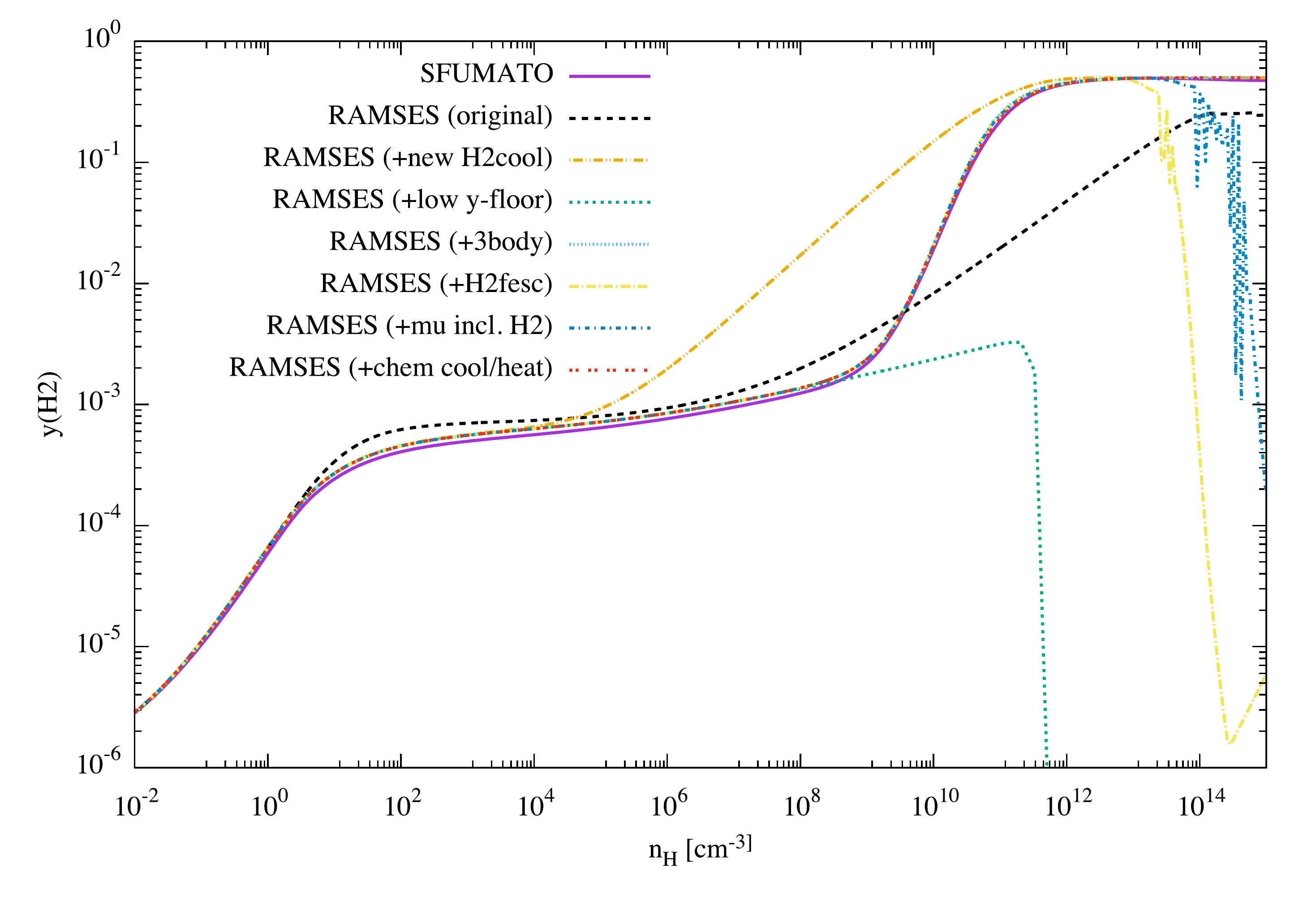}
	\includegraphics[width=0.48\textwidth]{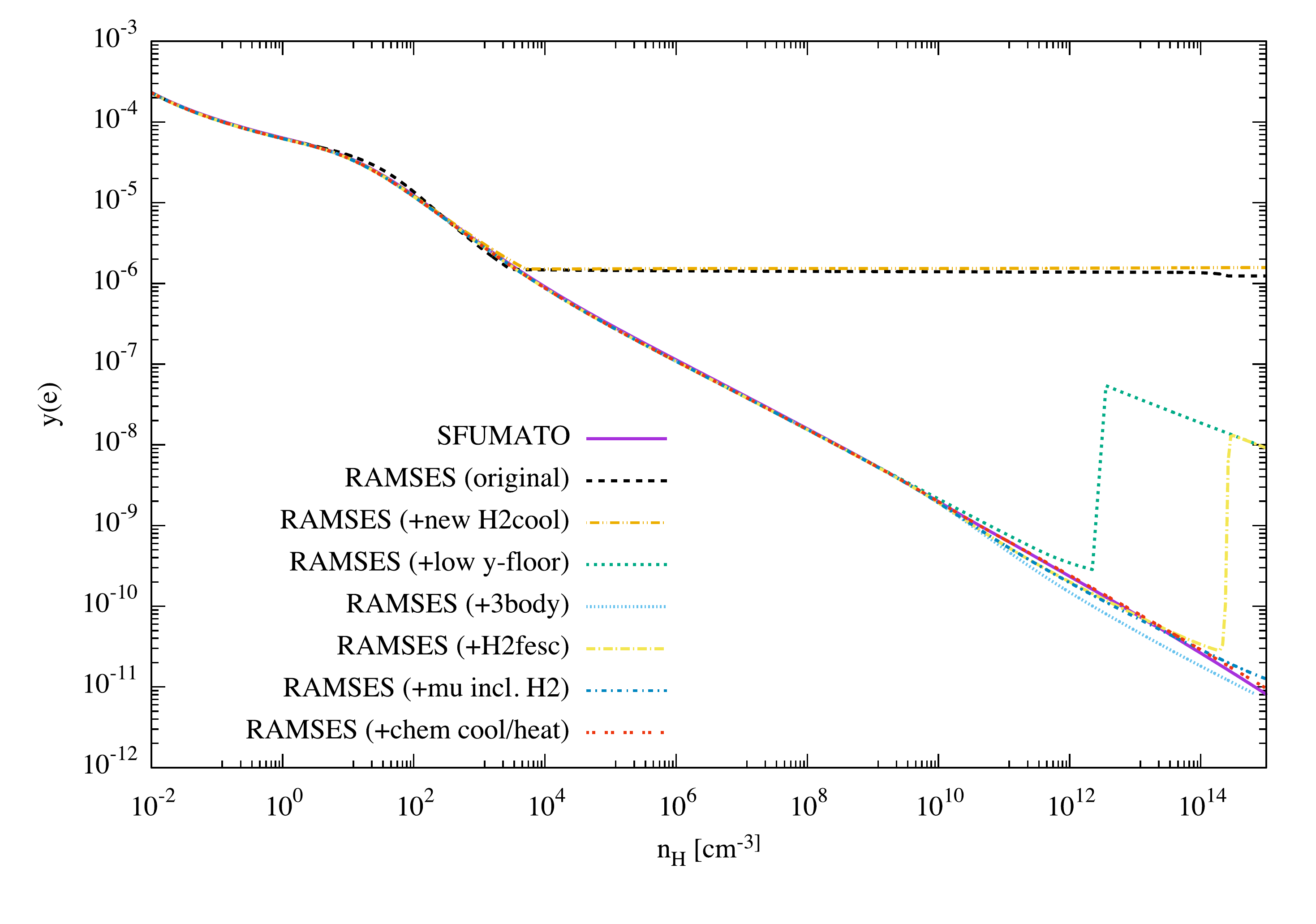}
    \caption{Same calculations as in Figure~\ref{fig:app1} but showing he molecular hydrogen abundance (left panel) and the electron fraction (right panel).}
    \label{fig:app2}
\end{figure*}

In Figure~\ref{fig:app1} we show the phase diagram for one-zone calculations (cooling balances compression heating,  \citep[e.g.][]{omukai2001} for successive improvements of the primordial chemistry/cooling module in \ramses. The different lines (see legend) illustrate how the original primordial cooling calculation in \ramses\ is improved with the addition of each of the physical processes discussed in Section~\ref{sec:sim}. We also compare the results with that of a former work (S20), indicated by the solid line with label {\it "SFUMATO"}. Figure~\ref{fig:app2} shows the \htwo\ (left panel) and electron fraction (right panel) for the same one-zone calculation. One of the major differences in this work is produced by lowering the minimum floor for the ionization fraction. As can be seen in the right panel of Figure~\ref{fig:app2} the electron fraction is lower than $10^{-6}$ at $\nh\gtrsim 10^4$\hcc. In the original version, however, the floor is set to $10^{-6}$ and thus the electron fraction is always higher than this threshold. For this reason, the simulation (orange dashed line) shows the \htwo\ fraction abnormally higher than it is supposed to be (right panel of Figure~\ref{fig:app1}). In the current version of our code, the value of the floor is $10^{-12}$ and therefore the validity of the model is guaranteed up to $\nh\sim 10^{14}$\hcc. With three-body \htwo\ formation, the models are consistent with each other up to densities \nh$\sim 10^9$\hcc\ and other improvements of the physics in the high-density regime reduces the error up to $\lesssim 10^{14}$\hcc.



\bsp	
\label{lastpage}
\end{document}